\documentclass[aps,prb,reprint]{revtex4-1}
\usepackage{graphicx}
\usepackage{bm}
\usepackage{braket}
\usepackage[normalem]{ulem}
\usepackage{amsmath}
\usepackage{mathrsfs}
\usepackage{multirow}
\usepackage{rotating}
\usepackage{todonotes}
\usepackage{array}
\usepackage{color}
\usepackage{feynmf}
\usepackage[colorlinks=true,linkcolor=blue, citecolor=blue, urlcolor=blue]{hyperref}
\usepackage{color,soul}
\usepackage{mathtools}
\newcommand{\stkout}[1]{\ifmmode\text{\sout{\ensuremath{#1}}}\else\sout{#1}\fi}

\newcolumntype{L}[1]{>{\raggedright\let\newline\\\arraybackslash\hspace{0pt}}m{#1}}
\newcolumntype{C}[1]{>{\centering\let\newline\\\arraybackslash\hspace{0pt}}m{#1}}
\newcolumntype{R}[1]{>{\raggedleft\let\newline\\\arraybackslash\hspace{0pt}}m{#1}}

\newcommand{\mo}[1]{\textcolor{red}{#1}}

\renewcommand{\thispagestyle}[1]{}

\begin{document}
\title{Ab-initio tensorial electronic friction for molecules on metal surfaces: nonadiabatic vibrational relaxation }
\author{Reinhard J. Maurer}
\email[]{reinhard.maurer@yale.edu}
\affiliation{Department of Chemistry, Yale University, New Haven CT 06520, USA}
\author{Mikhail Askerka}
\affiliation{Department of Chemistry, Yale University, New Haven CT 06520, USA}
\author{Victor S. Batista}
\affiliation{Department of Chemistry, Yale University, New Haven CT 06520, USA}
\author{John C. Tully}
\affiliation{Department of Chemistry, Yale University, New Haven CT 06520, USA}

\begin{abstract}
Molecular adsorbates on metal surfaces exchange energy with substrate phonons and low-lying electron-hole pair excitations. In the limit of weak coupling, electron-hole pair excitations can be seen as exerting frictional forces on adsorbates that enhance energy transfer and  facilitate vibrational relaxation or hot-electron mediated chemistry. We have recently reported on the relevance of tensorial properties of electronic friction [Phys. Rev. Lett. \textbf{116}, 217601 (2016)] in dynamics at surfaces. Here we present the underlying implementation of tensorial electronic friction based on Kohn-Sham Density Functional Theory for condensed phase and cluster systems. Using local atomic-orbital basis sets, we calculate nonadiabatic coupling matrix elements and evaluate the full electronic friction tensor in the Markov limit. Our approach is numerically stable and robust as shown by a detailed convergence analysis. We furthermore benchmark the accuracy of our approach by calculation of vibrational relaxation rates and lifetimes for a number of diatomic molecules at metal surfaces. We find friction-induced mode-coupling between neighboring CO adsorbates on Cu(100) in a c(2x2) overlayer to be important to understand experimental findings.
\end{abstract}


\maketitle 

\section{Introduction}
\label{intro}

Energy transfer between molecules and surfaces can be facilitated by a number of competing channels,~\cite{Arnolds2011} including dissipation due to surface phonons, excitation of electron-hole pairs~\cite{Tully1993,Wodtke2004} (EHPs), and charge transfer. Which of these  mechanisms dominates energy transfer is dependent on many factors including vibrational frequencies,~\cite{Morin1992, Harris1990, Krishna2006, Forsblom2007, Shenvi2009, Bartels2013} intramolecular vibrational energy distribution and vibrational mode coupling,~\cite{Askerka2016, Uranga-Pina2014} surface coverage,~\cite{Omiya2014} and temperature.~\cite{Springer1994, Matsiev2011} The question of how much energy is transferred among these channels at what time scale is fundamental to gas-surface dynamics,~\cite{Wodtke2004, Shenvi2009, Bunermann2015, Novko2015} single molecule electronic devices,~\cite{Maurer2012, Wolf2009, Conklin2013} heat and electron transport,~\cite{Paulsson2008, Esposito2015, Gunst2016} hot-electron driven chemistry on surfaces,~\cite{Olsen2009, Park2015} and heterogeneous catalysis.~\cite{Aiga2013,Golibrzuch2015} Specifically for reactions at metal surfaces, energy transfer and coupling between electronic and vibrational excitations are relevant to many chemical processes such as desorption induced by electronic transitions (DIET)~\cite{Menzel1964, Guo1999} and EHP-induced relaxation of adsorbate vibrations.~\cite{Tremblay2010} The reason is the insufficient energy separation of electronic and vibrational excitations leading to a break-down of the Born-Oppenheimer approximation at metal surfaces.~\cite{Wodtke2004,Rahinov2011}

Complementary to many surface science techniques that are sensitive to vibrational energy flow, such as molecular surface scattering,~\cite{Cooper2012, Bunermann2015} transient infrared spectroscopy,~\cite{Morin1992,Owrutsky1992, Watanabe2010} inelastic electron tunneling spectroscopy,~\cite{Lorente2000, Kim2015} sum-frequency generation (SFG) spectroscopy,~\cite{Wang2015} and surface- and tip-enhanced Raman spectroscopy,~\cite{Klingsporn2014, Hartman2016} first principles molecular simulation is a key tool to provide valuable insights on chemical reaction dynamics at surfaces.~\cite{Golibrzuch2015} The main challenge for simulations of energy transfer between the adsorbate and the substrate is to properly account for nonadiabatic couplings and their influence on molecular dynamics (MD).
The inability to apply fully quantum dynamics simulations to large systems has triggered the development of a wide range of quantum-classical methods, including methods that describe nonadiabatic dynamics at metal surfaces using Ehrenfest dynamics,~\cite{Li2005, Lindenblatt2006, Grotemeyer2014} surface-hopping,~\cite{Tully1990, Shenvi2009, Shenvi2009a} or approaches based on chemical master equations.~\cite{Dou2015,Dou2016,Dou2016a} 

An approach that is widely used~\cite{Tully1993,Kindt1998,Forsblom2007,Juaristi2008, Blanco-Rey2014,Rittmeyer2015,Novko2015,Loncaric2016} due to its simplicity and computational efficiency is molecular dynamics with electronic friction (MDEF).~\cite{DAgliano1975, Head-Gordon1995} MDEF describes the effect of electronic excitations as frictional damping forces in a Langevin dynamics framework. Electronic friction is described according to Fermi's Golden rule, derived from first order time-dependent perturbation theory (TDPT).~\cite{Persson1982,Hellsing1984, Head-Gordon1995} Applications of MDEF have addressed the study of vibrational cooling mediated by EHP excitations.~\cite{Head-Gordon1992a,Forsblom2007, Krishna2006, Tremblay2010} In its most widespread form, MDEF has been applied by assuming a single isotropic atomic damping coefficient~\cite{DAgliano1975} defined as a function of the local electron density. This approach is termed local density friction approximation (LDFA).~\cite{Echenique1981,Echenique1986} LDFA has been successfully applied to describe the surface dynamics of adatoms~\cite{Li1992,Blanco-Rey2014} and diatomics,~\cite{Juaristi2008,Fuchsel2013, Blanco-Rey2014,Rittmeyer2015,Novko2015,Novko2016,Loncaric2016} despite some controversy in the case of molecular adsorbates.~\cite{Juaristi2008,Luntz2009,Juaristi2009}  We have recently assessed some of these approximations as compared to TDPT and found that the tensorial nature of electronic friction is essential to describe intramolecular vibrational energy distribution.~\cite{Askerka2016} Furthermore, recent work on tunneling junctions and general non-equilibrium systems has shown that tensorial friction can introduce non-conservative and magnetic field forces.~\cite{Lu2010,Lu2011,Lu2015}

Here we present an efficient implementation of electronic friction based on TDPT. TDPT is commonly applied to calculate vibronic coupling effects such as EHP-induced vibrational relaxation,~\cite{Persson1982,Hellsing1984,Head-Gordon1992,Head-Gordon1992a,Krishna2006, Forsblom2007} electronic lifetime broadening,~\cite{Hellsing2002} and transport properties.~\cite{Savrasov1996, Bauer1998} In our approach electronic friction is viewed as a tensorial quantity~\cite{Head-Gordon1995} that not only defines adsorbate energy loss, but also  energy transfer between adsorbate atoms. We review the derivation of this approach (Section ~\ref{theory}) and present a stable numerical realization and efficient implementation within Density Functional Theory that is equally applicable to condensed matter and finite cluster systems (see Sections~\ref{computational}, \ref{results-numerics}, and \ref{results-comparison}). The resulting approach addresses difficulties in previous formulations that arise from uncertainties in representing a continuous spectrum with discrete energy levels from electronic structure calculations. We validate the approach by calculating the vibrational lifetimes of a number of representative diatomics (see Section~\ref{results-diatomics}) and larger adsorbates on metal surfaces.~\cite{supplemental} Furthermore, we discuss electronic friction in the context of collective adsorbate overlayer motion (see section~\ref{results-c2x2}).

\section{Theory}
\label{theory}

\subsection{Electronic Friction and Vibrational Energy Loss}
\label{methods-lifetime-MDEF}

\begin{figure}
\centering\includegraphics[width=\columnwidth]{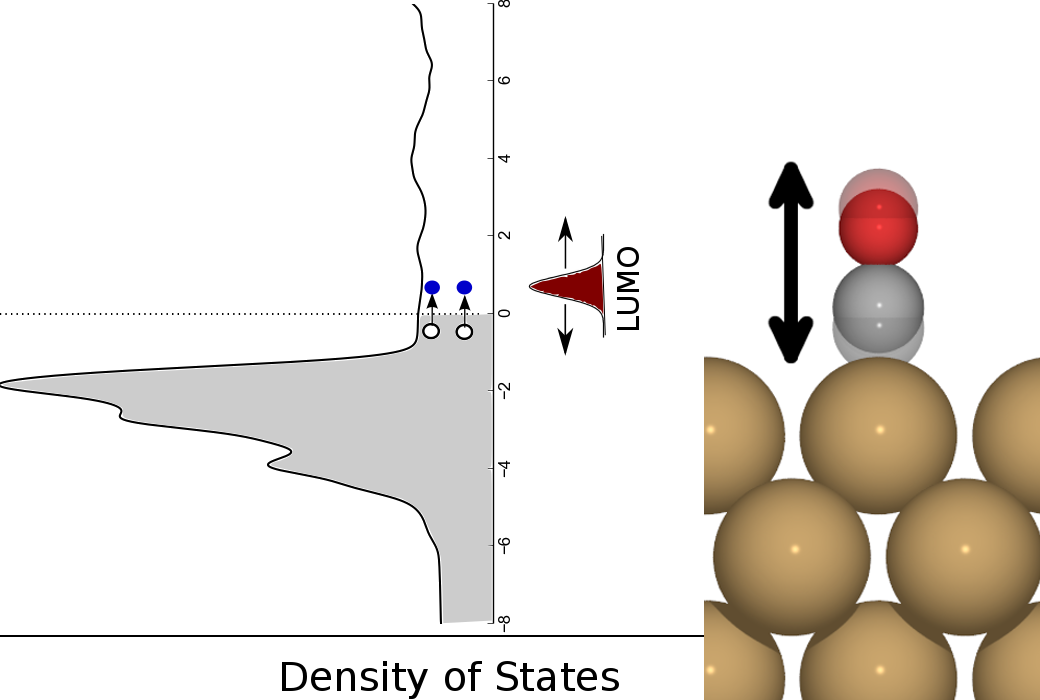}
\caption{\label{fig-intro} Vibrational motion of an adsorbate molecule (here shown is CO on a Cu(100) surface) induces low-lying electronic excitations. This leads to charge fluctuations and energy flow between substrate and adsorbate degrees of freedom. Especially for charge fluctuations that occur in molecular electronic states such as the lowest unoccupied molecular orbital (LUMO), nonadiabatic energy transfer can be large.}
\end{figure}

The vibrational and electronic spectra of most semi-conductors and insulating materials are usually well separated. At metal surfaces, however, the electronic and vibrational spectra overlap due to the vanishing energy gap between occupied and unoccupied energy levels, so even the smallest vibrational motions are coupled to EHP excitations (see Fig. \ref{fig-intro}). The underlying vibronic coupling affects the properties of the metal that relate to conduction, heat transport, and the transition to the superconducting state. Several studies have shown that the energy transfer between EHPs and the adsorbate can significantly alter the dynamics of surface scattering and surface reactions,~\cite{Blanco-Rey2014} for example by reducing transient mobility upon dissociation.~\cite{Kindt1998} Such nonadiabatic couplings are essential for a qualitatively correct description of surface scattering events~\cite{Shenvi2009a, Bunermann2015} and hot-electron induced reactions.~\cite{Springer1994} 

The MDEF method introduced by Head-Gordon and Tully~\cite{Head-Gordon1995} allows to incorporate nonadiabatic energy loss due to EHPs into classical MD simulations. The method is derived by transformation of mixed-quantum-classical equations of motion for classical nuclei and a quantum mechanical electron dynamics into action-angle variables~\cite{Miller1978,Meyer1979} and subsequently a generalized Langevin replacement~\cite{Adelman1976,Tully1980} of the explicit electronic degrees of freedom. Upon invoking the Markov or quasi-static approximation, we arrive at a classical Langevin equation of motion for the nuclear degree of freedom $i$ that is valid in the limit of weak coupling:
\begin{align}\label{eq-langevin}
M\ddot{R_i}= - \frac{\partial{V\bm{(R)}}}{\partial{R_i}} - \sum_{j} \Lambda_{ij}\dot{R_j} + \mathscr{R}_i(t).
\end{align}
The first term on the right-hand side of eq.~\ref{eq-langevin} corresponds to classical conservative dynamics as given by an adiabatic potential energy surface $V$($\mathbf{R}$), whereas the second term is the nonadiabatic friction term featuring the (3$\times$3N) electronic friction tensor $\mathbf{\Lambda}$ where N is the number of atoms, typically restricted to the adsorbate atoms. The third term is due to detailed balance and the fluctuation-dissipation theorem.~\cite{Callen1951} It corresponds to a statistical random force that satisfies:
\begin{equation}\label{eq-randomforce}
 \braket{\mathscr{R}_i(t)\mathscr{R}_j(0)} = k_B T\cdot\Lambda_{ij} . 
\end{equation}

The elements $\Lambda_{ij}$, in Eqs.~\ref{eq-langevin} and \ref{eq-randomforce}, are the usual friction tensor elements, describing the dragging force on coordinate $i$ due to motion along coordinate $j$. 
For equilibrium systems, $\mathbf{\Lambda}$ is strictly positive-definite, so friction on one atom always describes energy loss due to EHPs. Off-diagonal elements, however, can be positive or negative and thereby effectively describe energy flow between adsorbate atoms.~\cite{Askerka2016} In practical simulations, the off-diagonal elements of $\mathbf{\Lambda}$ in Cartesian space have often been assumed to be negligible.~\cite{Juaristi2008,Juaristi2009, Blanco-Rey2014} We have recently revisited this assumption for adatom and molecular motion on surfaces.~\cite{Askerka2016} Electronic friction exhibits a strong directional dependence and introduces couplings between vibrational degrees of freedom.

\begin{figure}
\centering\includegraphics[width=\columnwidth]{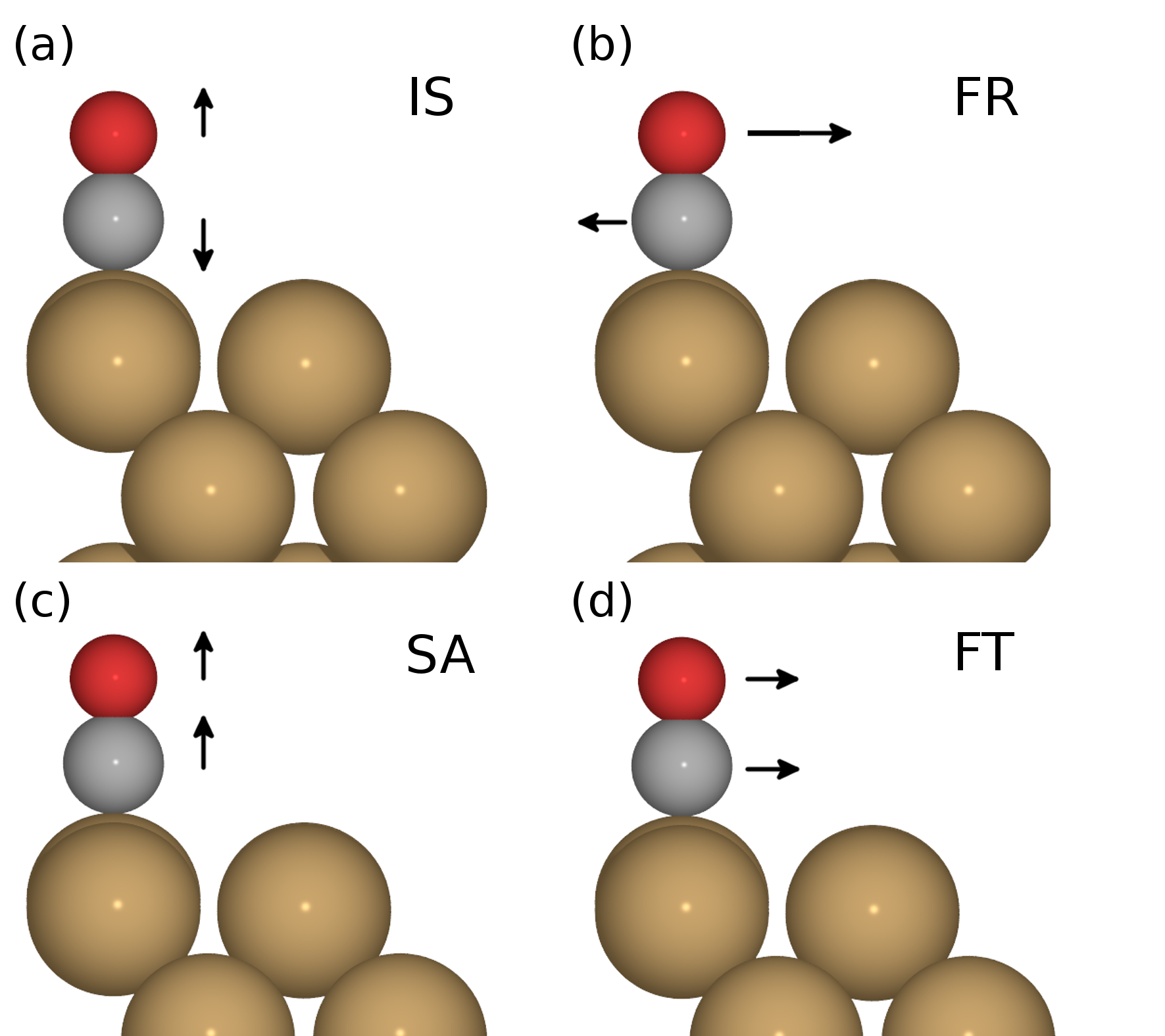}
\caption{\label{fig-vibmodes} Depiction of vibrational normal mode displacements of CO adsorbed at an atop site of a Cu(100) surface. (a) Internal stretch mode (IS), (b) Frustrated rotation mode (FR), (c) Surface-adsorbate mode (SA), and (d) Frustrated translation (FT) mode.}
\end{figure}

Molecular dynamics simulations based on eq.~\ref{eq-langevin} can describe vibrational relaxation. However, vibrational lifetimes can also
be estimated directly from the friction tensor assuming harmonic motion.~\cite{Head-Gordon1992,Head-Gordon1992a,Krishna2006,Forsblom2007} Vibrational lifetimes of molecular adsorbates can be measured by pump-probe spectroscopy.~\cite{Beckerle1991,Morin1992} 
In pump-probe experiments, a coherent pulse excites an adsorbate vibrational mode, \emph{e.g.} the internal stretch mode of CO adsorbed at Cu(100) in Fig.~\ref{fig-intro} (see also Fig.~\ref{fig-vibmodes}a). The vibrational frequency \mo{$\omega_j$} of this mode is well above the substrate phonon modes and its energy loss is thereby believed to be dominated by EHP excitations.~\cite{Morin1992,Omiya2014} Due to momentum conservation this mode is coherently excited in the full molecular overlayer. The corresponding collective, periodic motion can be described by a single adsorbate layer phonon mode with wavevector $\mathbf{q}=0$.
Assuming exponential decay, we can calculate the EHP-induced relaxation rate $\Gamma(\omega_j)$ of this vibration using the friction tensor of eq.~\ref{eq-langevin} by acknowledging the following relation:~\cite{Butler1979,Hellsing1984,Head-Gordon1992a}
\begin{align}\label{eq-vibrelax}
 \Gamma(\omega_j) =\mathbf{e}^T_{j}\cdot\mathbf{\tilde{\Lambda}}^{\mathbf{q=0}}\cdot\mathbf{e}_{j},
\end{align}
where $\mathbf{\tilde{\Lambda}}$ refers to the mass-weighted electronic friction tensor in Cartesian representation (or tensor of relaxation rates)  $\tilde{\Lambda}_{ij}=\Lambda_{ij}/(\sqrt{m_i}\sqrt{m_j})$ and $\mathbf{e}_{j}$ refers to the normalized displacement vector of the vibrational normal mode in question.

After laser excitation with frequency $\omega_j$, the coherently vibrating adsorbate overlayer will exhibit decoherence dephasing due to quasi-elastic scattering processes with low-lying substrate phonons and adsorbate vibrations and the motion of individual adsorbate molecules will decouple over time. The corresponding single adsorbate vibrational lifetime can be described by a totally symmetric linear combination of all possible phonons $\mathbf{q}$ at the perturbing frequency $\omega_j$:~\cite{Butler1979}
\begin{align}\label{eq-vibrelax2}
 & \Gamma(\omega_j) = \sum_{\mathbf{q}} w_{\mathbf{q}}\cdot \mathbf{e}^T_{\mathbf{q}j}\cdot\mathbf{\tilde{\Lambda}}^{\mathbf{q}}\cdot\mathbf{e}_{\mathbf{q}j} 
\end{align}
where $w_{\mathbf{q}}$ corresponds to a normalization weight. The friction tensor $\mathbf{\tilde{\Lambda}}^{\mathbf{q}}$ depends on $\mathbf{q}$ due to conservation rules between phonon and electron momenta. We will focus on vibrational relaxation of coherent adsorbate vibrations that can be described in simple surface unit cells and will only return to this topic in section~\ref{results-c2x2} of the results. The corresponding equations for the general $\mathbf{q}\neq0$ case are presented in the supplemental material.~\cite{supplemental}

The friction matrix $\mathbf{\Lambda}$ in Cartesian representation can be transformed into the normal mode representation by 
\begin{equation}\label{eq-normalmoderepresentation}
 \mathbf{\Lambda}_N = \mathbf{U} \mathbf{\Lambda}_C
\end{equation}
where $\mathbf{U}$ is the matrix of harmonic normal mode displacements $\mathbf{e}_j$ (eigenvectors of the mass-weighted vibrational Hessian matrix or dynamic matrix). As was recently shown,~\cite{Askerka2016} $\mathbf{\Lambda}$ is not necessarily diagonal in either of these representations. Therefore electronic friction as described by $\mathbf{\Lambda}_N$ induces couplings between normal modes.

\subsection{The Friction Tensor}
\label{methods-electron-phonon}

As described in the literature,\cite{Allen1972,Butler1979,Grimvall1981,Hellsing1984,Head-Gordon1992a,Quong1992,Savrasov1996,Onida2002, Trail2001,Mahan2013} the relaxation rate of a vibrational mode with frequency $\omega_{j}$ coupled to a manifold of energy levels described by Kohn-Sham (KS) DFT can be described by first order TDPT (see supplemental material for a derivation~\cite{supplemental}), as follows:
\begin{align}\label{eq-Gamma}
 \Gamma(\omega_{j}) &=\frac{\pi\hbar^2\omega_{j}}{M}\sum_{\mathbf{k},\nu,\nu'>\nu} 
    |g_{\mathbf{k}\nu,\nu'}^{j}|^2 \cdot \\ \nonumber & 
   [f(\epsilon_{\mathbf{k}\nu})-f(\epsilon_{\mathbf{k}\nu'})]\cdot \delta(\epsilon_{\mathbf{k}\nu'}-\epsilon_{\mathbf{k}\nu}-\hbar\omega_{j}),
\end{align}
where the sum over $\mathbf{k}$ indicates Brillouin sampling and implies normalization with corresponding weights w$_{\mathbf{k}}$. Electronic eigenstate energies $\epsilon$ are indicated with quantum numbers $\nu$,$\nu'$. For ease of presentation we have neglected spin, which introduces an additional sum over spin channels. The state occupation is given by the Fermi-Dirac distribution $f(\epsilon)$\footnote{The difference in populations arises from considering the probability of excitation and deexcitation at finite-temperature. When only considering deexcitations, the occupation factor would be $f(\epsilon_{\mathbf{k}\nu})\cdot(1-f(\epsilon_{\mathbf{k}\nu'}))$}, and the coupling elements between KS states $g_{\mathbf{k}\nu,\nu'}$ are defined as
\begin{align}\label{eq-g}
g_{\mathbf{k}\nu,\nu'}^{j} = \braket{\psi_{\mathbf{k}\nu}|\mathbf{e}_{j}\cdot\mathbf{\nabla}_{\mathbf{R}}|\psi_{\mathbf{k}\nu'}}.
\end{align}
The relaxation rate defines the lifetime of the mode $\tau=1/\Gamma$ and the vibrational linewidth $\gamma = \Gamma\hbar$. In eq.~\ref{eq-g}, $\mathbf{e}_{j}$ is the atomic displacement vector corresponding to the vibrational normal mode $\omega_{j}$ and $\mathbf{\nabla}_{\mathbf{R}}$ is the vector of Cartesian derivatives. 

We start by replacing the perturbing vibrational energy $\hbar\omega_j$ with the corresponding difference in electronic states:
\begin{align}\label{eq-Gamma2}
  \Gamma(\omega_{j}) &= \frac{\pi\hbar}{M} \sum_{\mathbf{k},\nu,\nu'>\nu} |g_{\mathbf{k}\nu,\nu'}^{j}|^2 \cdot (\epsilon_{\mathbf{k}\nu'}-\epsilon_{\mathbf{k}\nu}) \cdot \\ \nonumber &  [f(\epsilon_{\mathbf{k}\nu})-f(\epsilon_{\mathbf{k}\nu'})]\cdot\delta(\epsilon_{\mathbf{k}\nu'}-\epsilon_{\mathbf{k}\nu}-\hbar\omega_j). 
\end{align}
In the evaluation of eq.~\ref{eq-Gamma2}, the vibrational frequency $\omega_{j}$ only enters indirectly via the matrix elements and through the condition imposed by the delta-function. 
We limit our analysis to the so-called 'quasi-static' approximation, as described by, for example, Persson \emph{et al}~\cite{Hellsing1984}, which assumes that $\hbar\omega_j$ is much smaller than the energy at which the electron-phonon spectral density deviates from its low-energy asymptotic value.
It should be noted that there are two non-equivalent ways of doing this, both already mentioned in the seminal paper of Allen~\cite{Allen1972} and derived in appendix~\ref{appendix-alternative-expression}.

In order to obtain a convenient expression we now reformulate the relaxation rate by re-expressing the coupling elements $|g_{\mathbf{k}\nu,\nu'}^{j}|^2$ in terms of a Cartesian vibronic coupling tensor $\mathbf{A}$:
\begin{align}
 & |g_{\mathbf{k}\nu,\nu'}^{j}|^2 = g_{\mathbf{k}\nu,\nu'}^{j*}\cdot g_{\mathbf{k}\nu,\nu'}^{j} =\\ \nonumber
 & \mathbf{e}^T_{j}\cdot\braket{\psi_{\mathbf{k}\nu'}|\mathbf{\nabla}_{\mathbf{R}}|\psi_{\mathbf{k}\nu}}\cdot\braket{\psi_{\mathbf{k}\nu}|\mathbf{\nabla}_{\mathbf{R}}|\psi_{\mathbf{k}\nu'}} \cdot\mathbf{e}_{j}=  \\ \nonumber
 & \mathbf{e}^T_{j}\cdot \mathbf{A}_{\mathbf{k}\nu,\nu'}\cdot \mathbf{e}_{j} .
\end{align}
The elements of $\mathbf{A}$ are completely independent of vibrational modes and are defined as:
\begin{align}\label{eq-coupling}
 & A_{\mathbf{k}\nu,\nu'}^{n'a',na} = \braket{\psi_{\mathbf{k}\nu}|\frac{\partial}{\partial R_{n'a'}}|\psi_{\mathbf{k}\nu'}}\braket{\psi_{\mathbf{k}\nu'}|\frac{\partial}{\partial R_{na}}|\psi_{\mathbf{k}\nu}} ,
\end{align}
where $n$ and $n'$ indicate the $n$-th ($n'$-th) atom and $a$ and $a'$ indicate derivatives with respect to one of the three Cartesian directions. Inserting $\mathbf{A}$ into eq.~\ref{eq-Gamma2} we arrive at
\begin{align}
  &\Gamma(\omega_{j}) = \frac{\pi\hbar}{M} \sum_{\mathbf{k},\nu,\nu'>\nu} \mathbf{e}^T_{j}\cdot \mathbf{A}_{\mathbf{k}\nu,\nu'}\cdot \mathbf{e}_{j} \cdot (\epsilon_{\mathbf{k}\nu'}-\epsilon_{\mathbf{k}\nu}) \cdot \\ \nonumber & [f(\epsilon_{\mathbf{k}\nu})-f(\epsilon_{\mathbf{k}\nu'})]\cdot\delta(\epsilon_{\mathbf{k}\nu'}-\epsilon_{\mathbf{k}\nu}-\hbar\omega_j),
\end{align}
which we can further rewrite as
\begin{align}\label{eq-Gamma3}
  \Gamma(\omega_{j}) &= \frac{1}{M} \mathbf{e}^T_{j}\cdot  \biggl( \pi\hbar\sum_{\mathbf{k},\nu,\nu'>\nu} \mathbf{A}_{\mathbf{k}\nu,\nu'}\cdot (\epsilon_{\mathbf{k}\nu'}-\epsilon_{\mathbf{k}\nu}) \cdot  \\ \nonumber &    [f(\epsilon_{\mathbf{k}\nu})-f(\epsilon_{\mathbf{k}\nu'})]\cdot\delta(\epsilon_{\mathbf{k}\nu'}-\epsilon_{\mathbf{k}\nu}-\hbar\omega_j) \biggr) \cdot \mathbf{e}_{j} =\\ \nonumber &  \mathbf{e}^T_{j}\cdot\tilde{\mathbf{\Lambda}}(\omega_j)\cdot\mathbf{e}_{j}. 
\end{align}
The expression inbetween brackets, in eq.~\ref{eq-Gamma3}, corresponds to the Cartesian friction tensor $\mathbf{\Lambda}$ with dimensions $(3N\times3N)$ where N is the number of atoms considered. The elements of $\mathbf{\Lambda}$ are defined as
\begin{align}\label{eq-friction-tensor}
  \Lambda_{n'a',na}(\omega) &= \pi\hbar \sum_{\mathbf{k},\nu,\nu'>\nu} A_{\mathbf{k}\nu,\nu'}^{n'a',na} \cdot[f(\epsilon_{\mathbf{k}\nu})-f(\epsilon_{\mathbf{k}\nu'})] \cdot \\ \nonumber &(\epsilon_{\mathbf{k}\nu'}-\epsilon_{\mathbf{k}\nu})\cdot\delta(\epsilon_{\mathbf{k}\nu'}-\epsilon_{\mathbf{k}\nu}-\hbar\omega) .
\end{align}
Eq.~\ref{eq-Gamma3} and \ref{eq-Gamma} are equivalent since no further approximations have been made up to this point. When utilizing the friction tensor in the Langevin expression (eq.~\ref{eq-langevin}) or in harmonic relaxation rates (eq.~\ref{eq-vibrelax}) we will apply the quasi-static approximation (Appendix~\ref{appendix-alternative-expression}).

\subsection{Calculation of coupling and friction tensor}
\label{methods-coupling-matrix}

\emph{Ab-initio} evaluation of eq.~\ref{eq-friction-tensor} is numerically demanding. The two main computational challenges are the efficient evaluation of the nonadiabatic coupling matrix elements in $\mathbf{A}$ and the evaluation of the coupling strength in the quasi-static limit close to the Fermi level.

There are a number of ways to evaluate nonadiabatic coupling elements on the basis of Density Functional Theory.~\cite{Billeter2005,Fischer2011,Fatehi2011,Abad2013} Here, we employ ground-state KS states to calculate nonadiabatic couplings disregarding any final state effects or excited state screening. However, there may be simple ways to go beyond this approximation by employing effective excited-state screening methods such as constrained DFT,~\cite{Behler2007a,Carbogno2008} the Delta-self-consistent-field-DFT ($\Delta$SCF-DFT)~\cite{Ziegler1977} and related approaches,~\cite{Evangelista2013,Maurer2013} or the Slater transition potential method.~\cite{Slater1972,Triguero1998}

The matrix elements in eq.~\ref{eq-coupling} can be calculated from DFT using finite-difference schemes or analytically using Density Functional Perturbation Theory (Coupled-Perturbed Kohn-Sham).~\cite{Savrasov1996,Quong1992,Liu1996,Baroni2001} When using finite-difference schemes, specifically in a periodic plane-wave basis, problems due to arbitrary phase shifts can occur.~\cite{Lorente2000} For this purpose we have chosen to implement these expressions in a local atomic-orbital basis as implemented in SIESTA~\cite{Soler2002} and FHI-Aims.~\cite{Blum2009} This enables us to use the same framework for periodic and cluster calculations.

In a local atomic-orbital basis
\begin{equation}\label{eq-lcao}
\ket{\psi_{\mathbf{k}\nu}} = \sum_i c_{\mathbf{k}\nu}^i \ket{\phi_{\mathbf{k}}^i} ,
\end{equation}
the matrix elements of eq.~\ref{eq-coupling} can be reexpressed in terms of a generalized eigenvalue problem~\cite{Head-Gordon1992a} (see supplemental material for a more detailed derivation and the general $\mathbf{q}\neq0$ case~\cite{supplemental}):
\begin{align}\label{eq-HandS-pure}
 & \braket{\psi_{\mathbf{k}\nu}|\frac{\partial}{\partial R_{an}}|\psi_{\mathbf{k}\nu'}} = \\ \nonumber &\frac{  \mathbf{c}_{\mathbf{k}\nu'}\overbrace{\left(\mathbf{H}_{\mathbf{k}}^{an}-\epsilon_{\mathbf{k}\nu} \prescript{L}{}{\mathbf{S}}_{\mathbf{k}}^{an}-\epsilon_{\mathbf{k}\nu'} \prescript{R}{}{\mathbf{S}}_{\mathbf{k}}^{an} \right)}^{=\mathbf{G}} \mathbf{c}_{\mathbf{k}\nu}} {\epsilon_{\mathbf{k}\nu'}-\epsilon_{\mathbf{k}\nu}}
\end{align}
In eq.~\ref{eq-HandS-pure}, $\mathbf{c}_{\mathbf{k}\nu}$ represents the wavefunction expansion coefficient vector and $\mathbf{H}$, $\prescript{L}{}{\mathbf{S}}$, $\prescript{R}{}{\mathbf{S}}$ are the nuclear derivatives of the Hamiltonian and overlap matrices in the local atomic orbital representation:
\begin{equation}
 H^{an}_{\mathbf{k},ij} = \frac{\partial}{\partial R_{an}}\braket{\phi_{\mathbf{k}}^{i}|\hat{H}|\phi_{\mathbf{k}}^{j}} 
\end{equation}
and 
\begin{equation}
 {S}^{an}_{\mathbf{k},ij} = \prescript{L}{}{S}^{an}_{\mathbf{k},ij} + \prescript{R}{}{{S}}^{an}_{\mathbf{k},ij} = \braket{\frac{\partial}{\partial R_{an}}\phi_{\mathbf{k}}^i|\phi_{\mathbf{k}}^j} +  \braket{\phi_{\mathbf{k}}^i|\frac{\partial}{\partial R_{an}}\phi_{\mathbf{k}}^j}.
\end{equation}
We can evaluate the coupling matrix elements more conveniently by imposing certain approximations. For example by replacing the left- and right-sided derivative of the overlap matrix~\cite{Fatehi2011} with their average 
\begin{equation}
 \bar{S}^{an}_{\mathbf{k},ij} = \frac{1}{2} S^{an}_{\mathbf{k},ij} = \frac{1}{2} \frac{\partial}{\partial R_{an}}\braket{\phi_{\mathbf{k}}^i|\phi_{\mathbf{k}}^j} .
\end{equation}
The resulting nonadiabatic coupling elements are 
\begin{align}\label{eq-HandS-ave}
 & \braket{\psi_{\mathbf{k}\nu}|\frac{\partial}{\partial R_{an}}|\psi_{\mathbf{k}\nu'}} \approx \\ \nonumber &\frac{  \mathbf{c}_{\mathbf{k}\nu'}\overbrace{\left(\mathbf{H}_{\mathbf{k}}^{an}-\frac{1}{2}(\epsilon_{\mathbf{k}\nu} +\epsilon_{\mathbf{k}\nu'}) \mathbf{S}_{\mathbf{k}}^{an} \right)}^{=\mathbf{G}^{\mathrm{AVE}}} \mathbf{c}_{\mathbf{k}\nu}} {\epsilon_{\mathbf{k}\nu'}-\epsilon_{\mathbf{k}\nu}} .
\end{align}
We can furthermore eliminate the explicit occurance of eigenstate energies in the numerator by assuming that, on average, occupied and unoccupied states will be symmetric around the Fermi level $\epsilon_F\approx\frac{1}{2}(\epsilon_{\mathbf{k}\nu} +\epsilon_{\mathbf{k}\nu'})$:
\begin{align}\label{eq-HandS-HGT}
 & \braket{\psi_{\mathbf{k}\nu}|\frac{\partial}{\partial R_{an}}|\psi_{\mathbf{k}\nu'}} \approx \frac{  \mathbf{c}_{\mathbf{k}\nu'}\overbrace{\left(\mathbf{H}_{\mathbf{k}}^{an}-\epsilon_F \mathbf{S}_{\mathbf{k}}^{an} \right)}^{=\mathbf{G}^{\mathrm{HGT}}} \mathbf{c}_{\mathbf{k}\nu}} {\epsilon_{\mathbf{k}\nu'}-\epsilon_{\mathbf{k}\nu}} .
\end{align}
As described by \citet{Head-Gordon1992, Savrasov1996, Abad2013}, the matrix $\mathbf{G}^{\mathrm{HGT}}$ only depends on the converged wavefunction through the Hamiltonian and only requires derivatives of matrices in local basis representation. In the case of semi-empirical calculations, elements of $\mathbf{G}^{\mathrm{HGT}}$ can be precalculated and stored.~\cite{Abad2013} Section~\ref{results-comparison} analyzes the validity of approximations by comparing $\mathbf{G}$, $\mathbf{G}^{\mathrm{AVE}}$, and $\mathbf{G}^{\mathrm{HGT}}$.

Using the above coupling elements, the resulting vibronic coupling tensor is:
\begin{align}\label{eq-coupling-tensor-local}
  A_{\mathbf{k}\nu,\nu'}^{na,n'a'} = & \left(\frac{  \mathbf{c}^*_{\mathbf{k}\nu'}\mathbf{G}^{n'a'}_{\mathbf{k}} \mathbf{c}_{\mathbf{k}\nu}} {\epsilon_{\mathbf{k}\nu'}-\epsilon_{\mathbf{k}\nu}}\right)^*\cdot \frac{  \mathbf{c}^*_{\mathbf{k}\nu'}\mathbf{G}^{na}_{\mathbf{k}} \mathbf{c}_{\mathbf{k}\nu}} {\epsilon_{\mathbf{k}\nu'}-\epsilon_{\mathbf{k}\nu}}
\end{align}
Expressions of nuclear derivatives of $\mathbf{H}$ and $\mathbf{S}$ can be readily calculated
in existing electronic structure codes for couplings between electronic states with the same wavevector $\mathbf{k}$. In the supplemental material we describe  in detail how to calculate $\mathbf{G}_{\mathbf{k+q},\mathbf{k}}$ for $\mathbf{q}\neq 0$~\cite{supplemental}. For the remainder of this work we will confine ourselves to vibrational relaxation due to crystal periodic adsorbate motion ($\mathbf{q}=0$) or motion of an adsorbate in a cluster model. 

With an expression for the coupling tensor $\mathbf{A}$ at hand (eq.~\ref{eq-coupling-tensor-local}), what remains to be discussed is the evaluation of the friction tensor $\mathbf{\Lambda}(\omega)$ (eq.~\ref{eq-friction-tensor}). This is hindered by the slow convergence of the metal electronic structure with respect to Brillouin sampling close to the Fermi level, where the components of the friction tensor are to be evaluated in the quasi-static limit using a delta function.  Energy-resolved spectral properties such as the density of states (DOS), the friction tensor, and transport properties~\cite{Kawamura2014} are smooth functions in a periodic material and require highly accurate Brillouin zone integration and correspondingly a dense integration mesh in $\mathbf{k}$-space.~\cite{Monkhorst1976} Typically employed approximations to describe continuous manifolds of energy levels and to accelerate the convergence of Brillouin zone sampling include broadening methods~\cite{Methfessel1989} and different variants of the tetrahedron method.~\cite{Lehmann1972, Blochl1994, Kaprzyk2012,Kawamura2014} In the broadening method, a continuous integration over the first Brillouin zone is achieved by applying a broadening function to each point of an evenly distributed $\mathbf{k}$-point mesh. In order to achieve a smooth and stable DOS, in this work we replace the delta function in eq.\ref{eq-friction-tensor} with a Gaussian broadening function
\begin{equation}\label{eq-broadening}
 \delta(\epsilon_{i}-\epsilon_{j}) \approx \hat{\delta}(\epsilon_{i}-\epsilon_{j})  = \frac{1}{\sqrt{2\pi}\sigma}\cdot \exp \left\{\frac{-(\epsilon_{i}-\epsilon_{j})^2}{2\sigma^2}\right\} ,
\end{equation}
with a given broadening width $\sigma$. A number of different broadening functions are being used in the literature including Lorentzian, Hermite-Polynomials,~\cite{Methfessel1989} and 'squashed' Fermi-Dirac functions,~\cite{White1996,Trail2001} however we have found no discernible numerical advantage over one or the other. It was recently reported that spectral properties such as electron-phonon coupling sensitively depend on the choice of broadening $\sigma$ and significant errors in vibrational frequencies and vibrational damping rates can occur at large broadenings.~\cite{Kawamura2014} Our findings do not fully support this statement. We will address this point in detail in section~\ref{results-numerics}.

When calculating the components of the friction tensor in eq.~\ref{eq-friction-tensor}, the double sum over occupied and unoccupied states effectively runs over all electronic excitations starting from close to 0~eV (at the Fermi level $\epsilon_F$) up to high energies which barely contribute to the DOS at the Fermi level. At both limits of this energy range the broadening function $\hat{\delta}$ will be partly truncated and this 'leakage of excitation amplitude' needs to be corrected for by renormalization of the broadening function (eq.~\ref{eq-broadening}) according to the area over which the  couplings are acquired. We do this by dividing each excitation energy in eq.~\ref{eq-friction-tensor} with the analytical Gaussian integral from 0 to $\infty$:
\begin{equation}\label{eq-gaussian-broadening}
  \tilde{\delta}(\epsilon_{i}-\epsilon_{j}) =\frac{\hat{\delta}(\epsilon_{i}-\epsilon_{j}) }{\int_0^\infty \hat{\delta}(\epsilon-(\epsilon_i-\epsilon_j))\cdot d\epsilon} = \frac{\hat{\delta}(\epsilon_{i}-\epsilon_{j})}{\frac{1}{2}\left[1-\mathrm{erf}(\epsilon_i-\epsilon_j)\right]}.
\end{equation}
Our final working expression to evaluate friction in the quasi-static limit is
\begin{align}\label{eq-friction-final}
  &\Lambda_{n'a',na}(\omega) = \pi\hbar \cdot\\ \nonumber 
  &\sum_{\mathbf{k},\nu,\nu'>\nu} w_{\mathbf{k}}\frac{\left(  \mathbf{c}^*_{\mathbf{k}\nu'}\mathbf{G}^{n'a'}_{\mathbf{k}} \mathbf{c}_{\mathbf{k}\nu}\right)^*\cdot\mathbf{c}^*_{\mathbf{k}\nu'}\mathbf{G}^{na}_{\mathbf{k}} \mathbf{c}_{\mathbf{k}\nu}} {\epsilon_{\mathbf{k}\nu'}-\epsilon_{\mathbf{k}\nu}} \cdot \\ \nonumber & [f(\epsilon_{\mathbf{k}\nu})-f(\epsilon_{\mathbf{k}\nu'})]\cdot\tilde{\delta}(\epsilon_{\mathbf{k}\nu'}-\epsilon_{\mathbf{k}\nu}-\hbar\omega).
\end{align}
where we explicitly write the normalization weight associated with $\mathbf{k}$-grid integration. This expression differs from other nonadiabatic rate expressions in the literature that feature two separate delta functions for particle and hole excitations (eq.~\ref{eq-appendix-double-delta}), respectively.~\cite{Allen1972, Hellsing1984, Persson1982}. In appendix~\ref{appendix-alternative-expression} we show that, in the quasi-static limit, these two expressions are closely related.

\section{Computational details}
\label{computational}

We have implemented the above formalism in two forms. Our first implementation is a modular Python-based post-processing tool named \texttt{coolvib}~\cite{coolvib}. \texttt{coolvib} currently  wavefunctions, Hamiltonian, and overlap matrix output from two different local atomic-orbital electronic structure codes - the SIESTA~\cite{Soler2002} code using Gaussian basis functions and pseudoized core states and the FHI-Aims~\cite{Blum2009} code using all-electron numerical atomic orbitals. The second implementation corresponds to a direct, fully parallelized FORTRAN-based calculation of the electronic friction tensor within the current development version of FHI-Aims. In both cases, the calculation of the friction tensor is based on a previous calculation of KS eigenvalues and eigenvectors as well as the basis set representation of Hamiltonian and overlap. The input data can either be calculated analytically \emph{via} Density Functional Perturbation Theory (DFPT)~\cite{Savrasov1996,Baroni2001} or using finite difference methods. Due to the current (partial) lack of DFPT in both codes, we employ a symmetric finite-difference scheme where we displace the adsorbate atoms in all Cartesian directions. We chose a displacement of 10$^{-3}$\AA{}. However, our results do not vary significantly over a wide range of displacements (10$^{-2}$-10$^{-4}$~\AA{}). In section~\ref{results-comparison}, we employ analytical expressions for the left and right nuclear derivative of the overlap matrix as it is currently available in the FHI-Aims development version for cluster systems.

Relaxation rates and the friction tensor must be converged with respect to the substrate slab size, the basis set, Brillouin zone sampling mesh, and the maximum excitation energy cutoff. All of these parameters have been chosen to ensure convergence of vibrational lifetimes within a tolerance of 5-10\%. For our calculations, we have chosen a broadening width $\sigma$ in eq.~\ref{eq-broadening} of 0.6~eV. The electronic temperature in the friction expression was chosen as 300~K. It is found that the maximum excitation energy cutoff is safely converged at a value of $5\sigma$. More details on convergence and numerical stability are given in section~\ref{results-numerics}.

All calculations in sections~\ref{results-comparison}, \ref{results-diatomics}, and \ref{results-c2x2} have been performed using FHI-Aims and the Perdew, Burke, and Ernzerhof (PBE) exchange-correlation functional.~\cite{Perdew1996} The modeled metal slabs consist of 4 metal layers of which the lowest one was frozen during geometry optimizations. Atomic forces have been minimized down to a remaining residual force per atom of 0.01~eV/\AA{}. For all systems we used  optimized PBE bulk lattice constants and a vacuum above the slab that exceeded 50~{\AA}. Isolated metal clusters were constructed from bulk truncated surfaces and only the position of the adsorbate was optimized. Electronic structure convergence was set to residual changes per SCF cycle of $10^{-6}$ for the density and $10^{-6}$ eV for the total energy. Vibrational relaxation rates and lifetimes are calculated by transformation of the friction tensor to normal mode coordinates and selection of the corresponding diagonal element (see eq.~\ref{eq-vibrelax}). Vibrational modes were calculated using the finite-difference method.

\section{Results and Discussion}
\label{results}

In the following, we study in detail the numerical convergence properties of the DFT-based electronic friction tensor and the computational cost (section~\ref{results-numerics} and \ref{results-comparison}). We furthermore apply our approach to calculations of vibrational lifetimes for a number of diatomic molecules adsorbed on different metal surfaces (section~\ref{results-diatomics}) and study in detail intra- and intermolecular mode-coupling introduced by electronic friction for the example CO on Cu(100) (section~\ref{results-c2x2}). In the supplemental material we present calculations for a larger adsorbate, namely methoxide adsorbed on Cu(100).~\cite{supplemental} 

\subsection{Numerical considerations and convergence}
\label{results-numerics}

We study the numerical stability and convergence behavior of the friction tensor for calculations of vibrational lifetimes of p(2x2) CO adsorbed at the atop site of a bulk-truncated Cu(100) surface (see Fig.~\ref{fig-vibmodes}). We do this by computing the relaxation rates of adsorbate normal modes. This requires transformation of the friction tensor to normal mode coordinates and selection of the corresponding diagonal element. These calculations have been performed using the SIESTA code~\cite{Soler2002} and the PBE functional.~\cite{Perdew1996} A plane wave cut-off of 200 Ry was used for the grid, the forces were converged at 0.004 eV/\AA{}, the SCF equations were converged at residual energies of $10^{-5}$ eV with the electronic temperature of 290~K to facilitate the SCF convergence for the metallic slab. In the convergence run, 3 layers of Cu atoms were used; all of the metal atoms were kept frozen during the optimization. For these convergence tests, electronic friction was calculated at a temperature of 1000~K.

\begin{figure}
\centering\includegraphics[width=\columnwidth]{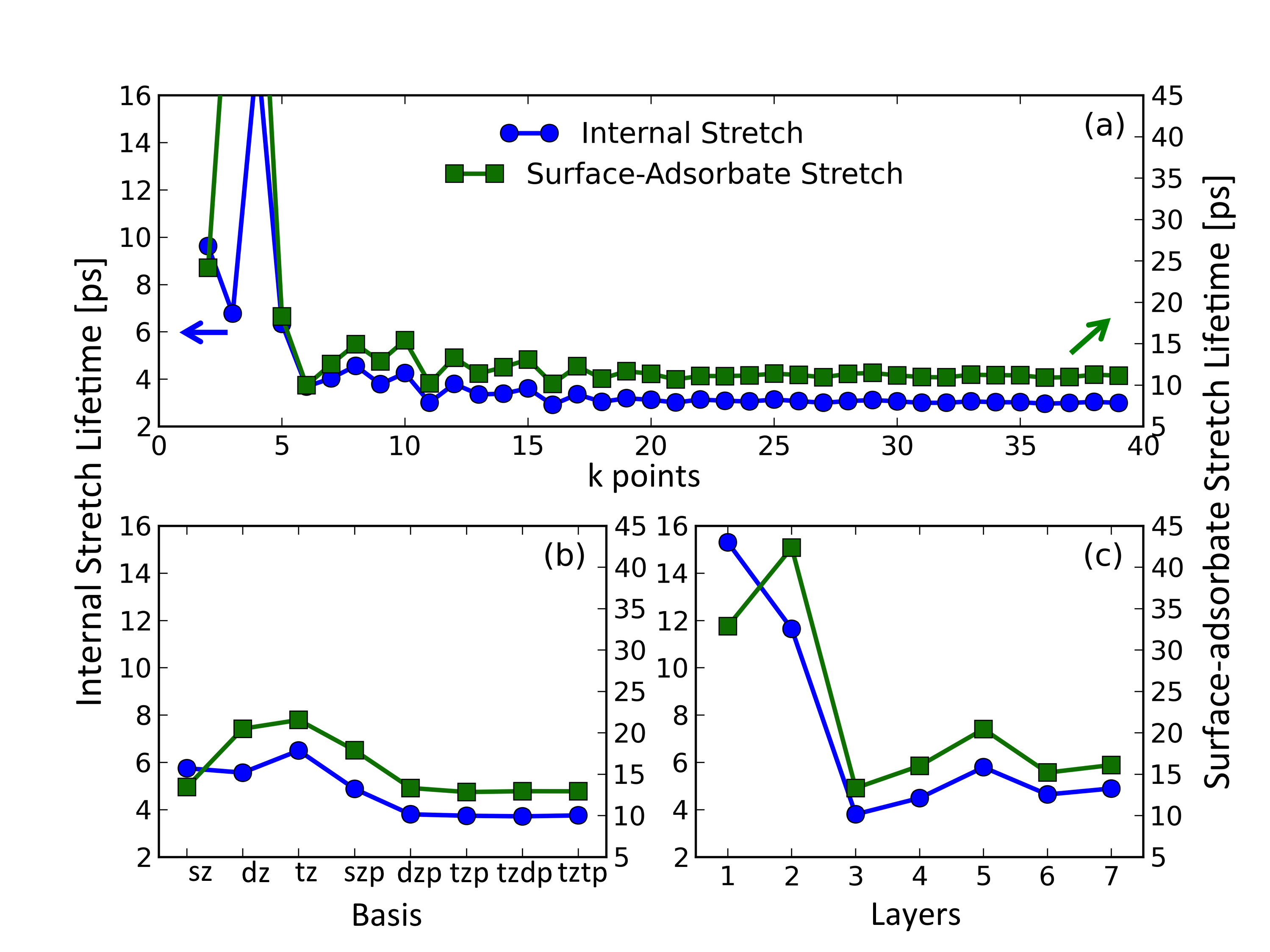}
\caption{\label{fig-convergence} Projected lifetimes of atop CO on Cu(100) along the IS and the SA mode (see Fig~\ref{fig-vibmodes}). (a) Convergence with respect to $a\times a\times a$ $\mathbf{k}$-grid, with $a$ being the labelling on the x-axis. (b) Convergence with respect to basis set: sz, dz, and tz correspond to single, double, and triple valence; sp, dp, and tp correspond to single, double, and triple polarization functions. (c) Convergence with respect to metal layers in the substrate. All lifetimes are given in picoseconds.}
\end{figure}

Fig.~\ref{fig-convergence} shows the projected vibrational lifetimes along approximate vectors $\vec{v}_{IS}=(-\sqrt{1/2},\sqrt{1/2})$ and $\vec{v}_{SA}=(\sqrt{1/2},\sqrt{1/2})$ perpendicular to the surface. 
The convergence with respect to Brillouin zone sampling and the employed Monkhorst-Pack $\mathbf{k}$-grid~\cite{Monkhorst1976} is rapid and convergence within 5\% variation of the lifetime is reached at a moderately dense $18\times18\times1$ $\mathbf{k}$-grid amounting to 162 unique $\mathbf{k}$-points for both modes. In fact, we find the relative convergence of relaxation rates to be about the same for all elements of the friction tensor. The $\mathbf{k}$-convergence is also related to the broadening width $\sigma$ in eq.~\ref{eq-friction-final}, which we will discuss further below. 

Convergence with respect to the basis set size has been fast for all studied cases. As shown in Fig.~\ref{fig-convergence}, vibrational lifetimes can be considered converged already at a double valence basis set with additional polarization functions when using Gaussian basis functions. In the case of numerical atomic orbitals as employed in FHI-Aims, we find convergence using 'tight' integration settings and a Tier 2 basis set corresponding to two full valence shells of basis functions per free atom state. The convergence with respect to the number of substrate layers is slower with a remaining change of the IS lifetime from 4.5 to 4.8~ps between 4 and 7 substrate layers. We find again that the convergence behavior is consistent for both modes. Considering the underlying approximations in the calculation of the electronic friction tensor, we consider the convergence tolerance of below 10\% that can be achieved with 4 substrate layers to be sufficiently accurate for our purposes.

\begin{figure}
\centering\includegraphics[width=\columnwidth]{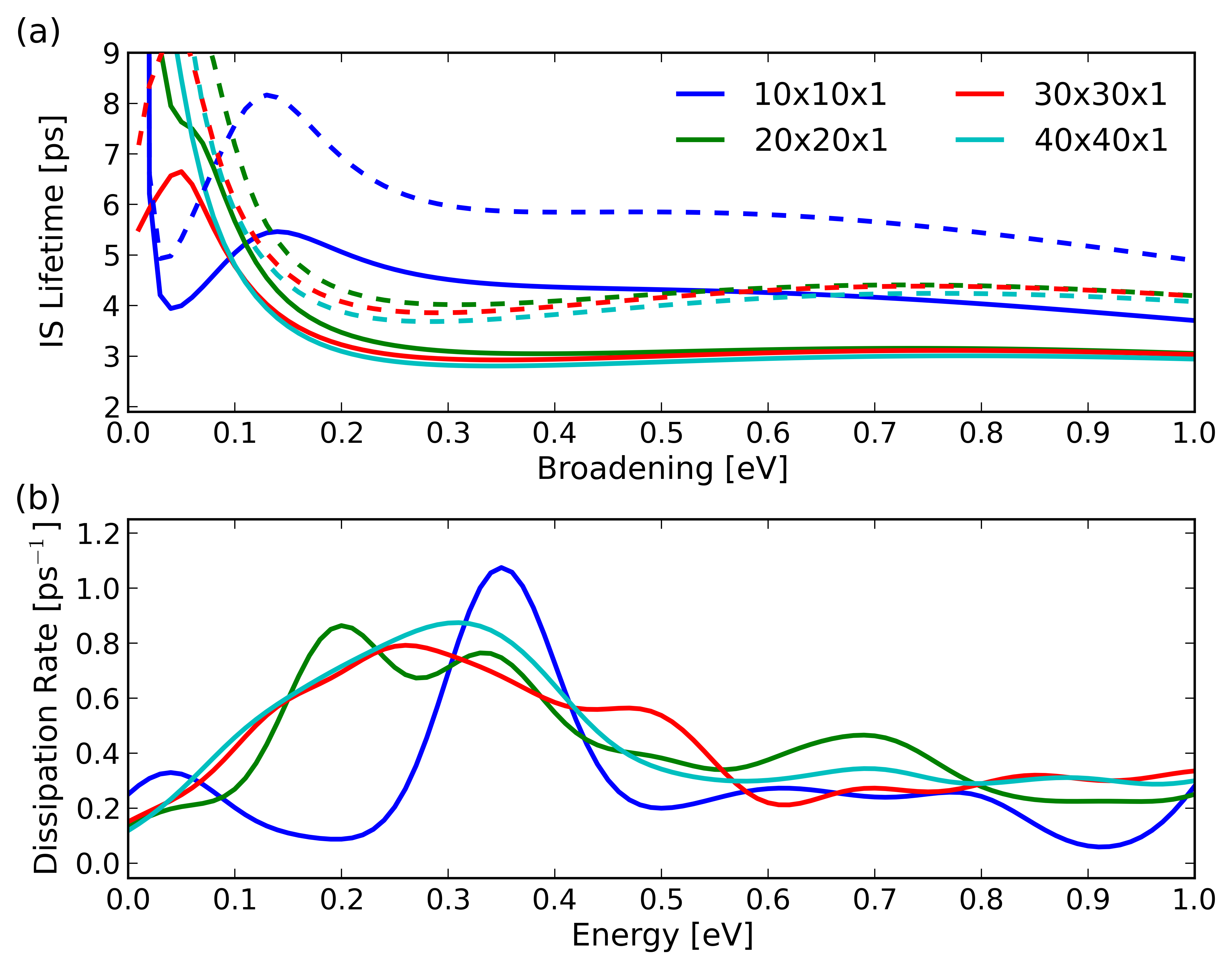}
\caption{\label{fig-convergence2} (a) Dependence of the IS lifetime of CO on Cu(100) on the delta-function broadening $\sigma$ for various k-point sets with (solid) and without (dashed) delta function normalization as described in eq.~\ref{eq-gaussian-broadening}. (b) Dissipation rate along the IS mode as a function of $\hbar\omega$ (eq.\ref{eq-friction-final}); the Gaussian broadening of 0.05 eV is used to smoothen the spectrum.}
\end{figure}

We now turn to the dependence of relaxation rates and vibrational lifetimes with respect to the parameter $\sigma$ that enters eq.~\ref{eq-friction-final} through the Gaussian broadening function. Fig.~\ref{fig-convergence2}a shows the IS lifetime of CO on Cu(100) as a function of broadening width $\sigma$. Broadening facilitates Brillouin zone integration and should be chosen large enough to result in a converged DOS at the Fermi level, but small enough not to bias the relaxation rate.~\cite{Methfessel1989,Pickard1999} At small broadening widths we find the lifetime to change rapidly up to a point at around 0.3~eV after which it remains almost constant up to 1~eV. This behavior can be understood by inspection of the corresponding projected element of the friction tensor as a function of $\hbar\omega$ (Fig.~\ref{fig-convergence2}b). We limit our analysis to the evaluation of $\Lambda(0)$. By increasing $\sigma$ we smoothen the function and introduce higher-energy contributions. This effect leads to an initially large change of lifetime at small broadenings due to the large spectral intensity around 0.05~eV. However, for higher broadening values the lifetime becomes almost independent of $\sigma$. It is interesting to note that we can clearly distinguish qualitatively unconverged from converged Brillouin sampling, simply by comparing the broadening dependence in Fig.~\ref{fig-convergence2}a  with $\Lambda(\omega)$ for different $\mathbf{k}$-grids (Fig.~\ref{fig-convergence2}b, blue solid curve). At and above a fully converged $\mathbf{k}$-grid the lifetime is largely independent of broadening width and Brillouin sampling. 

\citet{Kawamura2014} have recently argued that vibrational frequencies and electron-phonon linewidths calculated with broadening-based methods strongly depend on the broadening width and significant deviations may be introduced when compared to interpolation methods. The authors find a strong variation in the McMillan onset temperature of superconductivity at broadenings ranging from 0.1 to 0.7~eV with no sign of convergence.  
We find correct normalization of the broadening function to be critical to achieve stability over a wide range of broadening widths. Broadening functions that are not normalized over the correct domain, in our case (0,$\infty$], lead to a larger systematic drift in lifetime at larger values of $\sigma$ (see dashed lines in Fig.~\ref{fig-convergence2}(a)). 


\begin{figure}[h!]
\centering\includegraphics[width=\columnwidth]{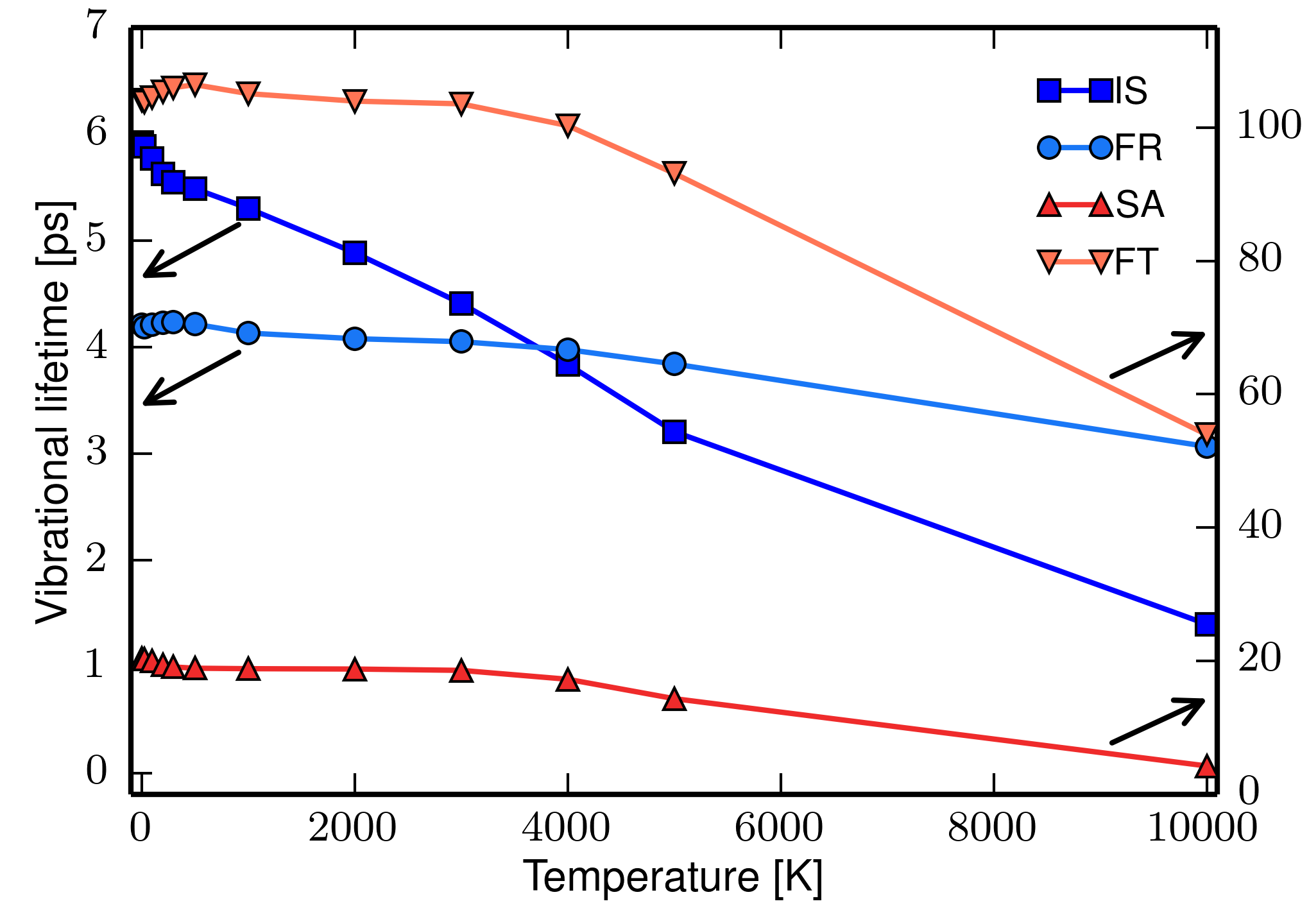}
\caption{\label{fig-temperature} Vibrational lifetimes as a function of electronic temperature due to electron-hole pairs for 4 different vibrational modes of CO adsorbed atop on Cu(100). The vibrational modes are abbreviated as in Fig.~\ref{fig-vibmodes}. }
\end{figure}

Finally, we address the electronic temperature that enters the state populations in eq.~\ref{eq-friction-final}. The elements of the friction tensor and therewith calculated vibrational lifetimes do not show a strong dependence on electronic temperature at low to ambient temperatures. For CO on Cu(100) we find lifetimes to vary only up to 10\% in a window of 0 to 2000~K (see Fig.~\ref{fig-temperature}). At low to intermediate electronic temperatures between 300 and 2000~K we find only small temperature-dependence of the vibrational lifetimes with the exception of the internal stretch (IS) mode. The IS mode lifetime changes between 6 and 5~ps in this electronic temperature regime. For laser-driven excitations, electronic temperatures of many thousands of Kelvin for a short period of time are not unusual. At such higher temperatures, vibrational lifetimes can change up to an order of magnitude. Above 4000~K (or $k_bT$ of 0.34~eV) we find a significant reduction of lifetime (or increase in relaxation rate). The effect of finite electronic temperature in eq.~\ref{eq-friction-final} is to mix higher-energy contributions of the spectral function into $\Lambda(0)$.  Depending on the shape of the spectral function along a certain motion this can initially lead to reduction or increase of relaxation rates. At high electronic temperatures this leads to a reduction of vibrational lifetime, simply due to the larger average spectral density above and below the Fermi level (compare also with Fig.~\ref{fig-convergence2}b). In all further calculations below we have chosen an electronic temperature of 300~K.

\subsection{Comparison of different coupling expressions}
\label{results-comparison}

\begin{figure}
\centering\includegraphics[width=\columnwidth]{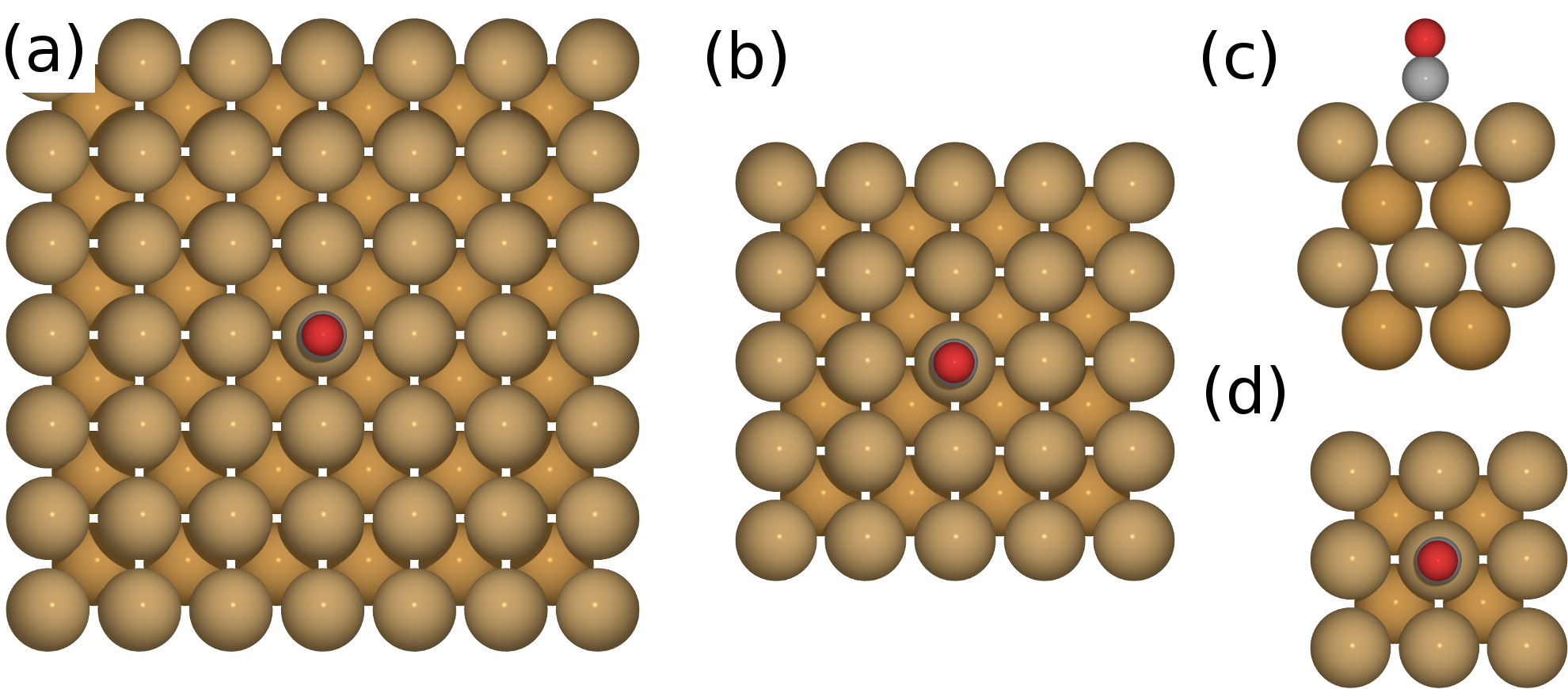}
\caption{\label{fig-cluster-structure} CO adsorbed on (100) bulk truncated Cu clusters of size (7x7) (a), (5x5) (b), and (3x3) in side (c) and top (d) view. Odd and even layers of substrate are shown in different color tones.}
\end{figure}

In this section, we study the validity of different approximations to the nonadiabatic coupling matrix elements $\mathbf{G}$ that enter the expression for the friction tensor. In section~\ref{methods-coupling-matrix} we have identified three different expressions for the coupling matrix elements: (1) the exact coupling elements $\mathbf{G}$ (eq.~\ref{eq-HandS-pure}), which require the evaluation of left- and right-sided nuclear derivatives of the overlap matrix and KS energies explicitly enter $\mathbf{G}$, (2) coupling elements $\mathbf{G}^{\mathrm{AVE}}$ (eq.~\ref{eq-HandS-ave}), where we replace $\prescript{L}{}{\mathbf{S}}$ and $\prescript{R}{}{\mathbf{S}}$ with an average $\bar{\mathbf{S}}$ corresponding to a symmetric matrix derivative of the overlap, and (3) coupling elements $\mathbf{G}^{\mathrm{HGT}}$ (eq.~\ref{eq-HandS-HGT}), where we replace the explicit KS eigenenergies by the Fermi energy. Coupling elements $\mathbf{G}^{\mathrm{HGT}}$ are arguably the most convenient, because they do not explicitly depend on the KS wavefunctions and eigenenergies and can be evaluated once for every geometry, independent of $T$ or $\sigma$.~\cite{Head-Gordon1992} This is especially useful for semi-empirical methods.~\cite{Abad2013} We have performed vibrational lifetime calculations using all three expressions for CO adsorbed on bulk-truncated Cu(100) clusters of different sizes (see Fig.~\ref{fig-cluster-structure}). Each cluster contains 4 layers of substrate and different lateral dimensions corresponding to (3x3), (5x5), and (7x7) Cu(100) unit cells.

\begin{table}
\caption{\label{tab-comparison} Comparison of different expressions for the nonadiabatic coupling elements entering the friction tensor, namely $\mathbf{G}$ (eq.~\ref{eq-HandS-pure}), $\mathbf{G}^{\mathrm{AVE}}$ (eq.~\ref{eq-HandS-ave}), and $\mathbf{G}^{\mathrm{HGT}}$ (eq.~\ref{eq-HandS-HGT}) for CO adsorbed on (100)-truncated Cu clusters of different size (see Fig.~\ref{fig-cluster-structure}). Shown are diagonal ($C_{zz}$, and $O_{zz}$) and off-diagonal mass-weighted friction tensor elements (C$_z$O$_{z}$) perpendicular to the surface as well as vibrational lifetimes projected along the internal stretch (IS) and surface-adsorbate (SA) modes.  }
\begin{tabular}{ccC{1.0cm}C{1.0cm}C{1.4cm}C{1.0cm}C{1.0cm}} \hline \rule{0pt}{2ex}
system & couplings & $C_{zz}$  & $O_{zz}$ & C$_z$O$_{z}$ & IS  & SA  \\ 
   &     & \multicolumn{3}{c}{rates in ps$^{-1}$} & \multicolumn{2}{c}{lifetimes in ps} \\ \hline \noalign{\vskip 2pt} 
(3x3)         & $\mathbf{G}^{\mathrm{HGT}}$  & 0.157 & 0.147 & -0.138 & 3.47 & 64.7    \\ 
cluster       & $\mathbf{G}^{\mathrm{AVE}}$  & 0.180 & 0.159 & -0.154 & 3.10 & 63.1    \\
              & $\mathbf{G}$ & 0.189 & 0.159 & -0.158 & 3.01 & 64.0    \\
(5x5)         & $\mathbf{G}^{\mathrm{HGT}}$  & 0.189 & 0.056 & -0.088 & 4.53 & 41.8     \\ 
cluster       & $\mathbf{G}^{\mathrm{AVE}}$  & 0.192 & 0.057 & -0.089 & 4.46 & 41.3     \\
              & $\mathbf{G}$ & 0.197 & 0.057 & -0.090 & 4.38 & 39.8     \\
(7x7)         & $\mathbf{G}^{\mathrm{HGT}}$  & 0.323 & 0.090 & -0.126 & 2.85 &  16.3    \\ 
cluster       & $\mathbf{G}^{\mathrm{AVE}}$  & 0.327 & 0.092 & -0.130 & 2.79 &  16.5    \\
              & $\mathbf{G}$ & 0.334  & 0.091 & -0.131 & 2.75 & 16.1     \\ \hline
\end{tabular} 
\end{table}

The convergence of elements of the electronic friction tensor and of the projected vibrational lifetimes with respect to cluster size is slow (Table~\ref{tab-comparison}) and independent of the applied nonadiabatic coupling elements. This is particularly true for the surface-adsorbate stretch mode (SA). As mentioned before, we have recently found that off-diagonal elements can be negative and of comparable size as diagonal elements. By inspection of the individual Cartesian elements perpendicular to the surface, we find that this also holds for CO adsorbed on Cu clusters of different sizes. The off-diagonal elements describe rate transfer between Cartesian directions and cannot be neglected. For example, the projected IS and SA lifetimes of CO adsorbed at the (7x7) cluster without off-diagonal elements would be 4.7~ps rather than 2.75~ps and 16.1~ps. A positive off-diagonal would reduce the lifetime in the SA and increase the lifetime in the IS mode. We calculate a negative off-diagonal component of -0.131~ps$^{-1}$ that reduces the lifetime of the IS and increases the lifetime of the SA mode.

When replacing the correct overlap derivatives by the average of the two ($\mathbf{G}^{\mathrm{AVE}}$), we find similar relaxation rates and lifetimes with deviations well within our numerical convergence tolerance. This approximation, in the studied systems, seems to lead to a minor reduction of diagonal and off-diagonal relaxation rates. The resulting effect on the vibrational lifetimes can be both an over- or underestimation, depending on the mode. When additionally introducing the assumption that excitations are close enough to the Fermi level and symmetric, we arrive at the original coupling expressions of Head-Gordon and Tully ($\mathbf{G}^{\mathrm{HGT}}$).~\cite{Head-Gordon1992} The overall change in relaxation rates and vibrational lifetimes is again minor and within numerical tolerances. However, two points related to the two approximations deserve some closer attention. First of all, both approximations have a stronger impact on the (3x3) cluster than on the larger ones. This cluster has a finite gap (0.04~eV at the PBE level) and in this system the DOS is far from being a continuous spectrum. Correspondingly, the assumption that occupied and unoccupied states will be equally spaced around the Fermi level will be less justified than in the case of a continuous and constant DOS around the Fermi level. This effect might also be more pronounced for transition metal systems with open d-shells, such as Nickel and Cobalt. Secondly, the latter approximation leads to a small but systematic reduction of relaxation rates, giving an overestimation of vibrational lifetimes. 

In summary, we find the approximations taken by Head-Gordon and Tully and others~\cite{Head-Gordon1992, Abad2013} to be valid for the tested systems. However, for future work we may consider using the exact coupling elements, since in KS-DFT calculations the difference in computational expense is negligible. In the case of semi-empirical electronic structure calculations such as performed using the FIREBALL method,~\cite{Sankey1989, Lewis2001, Abad2013} the above approximations are crucial to enable a tabulation of matrix derivatives prior to calculation and an efficient evaluation of nonadiabatic couplings. For the remainder of this work we will apply $\mathbf{G}^{\mathrm{HGT}}$ (eq.~\ref{eq-HandS-HGT}) to calculate energy dissipation in example systems such as diatomic and multiatomic molecules adsorbed on metal surfaces.

\subsection{Vibrational relaxation of diatomic molecules on metal surfaces}
\label{results-diatomics}

\begin{figure}
\centering\includegraphics[width=\columnwidth]{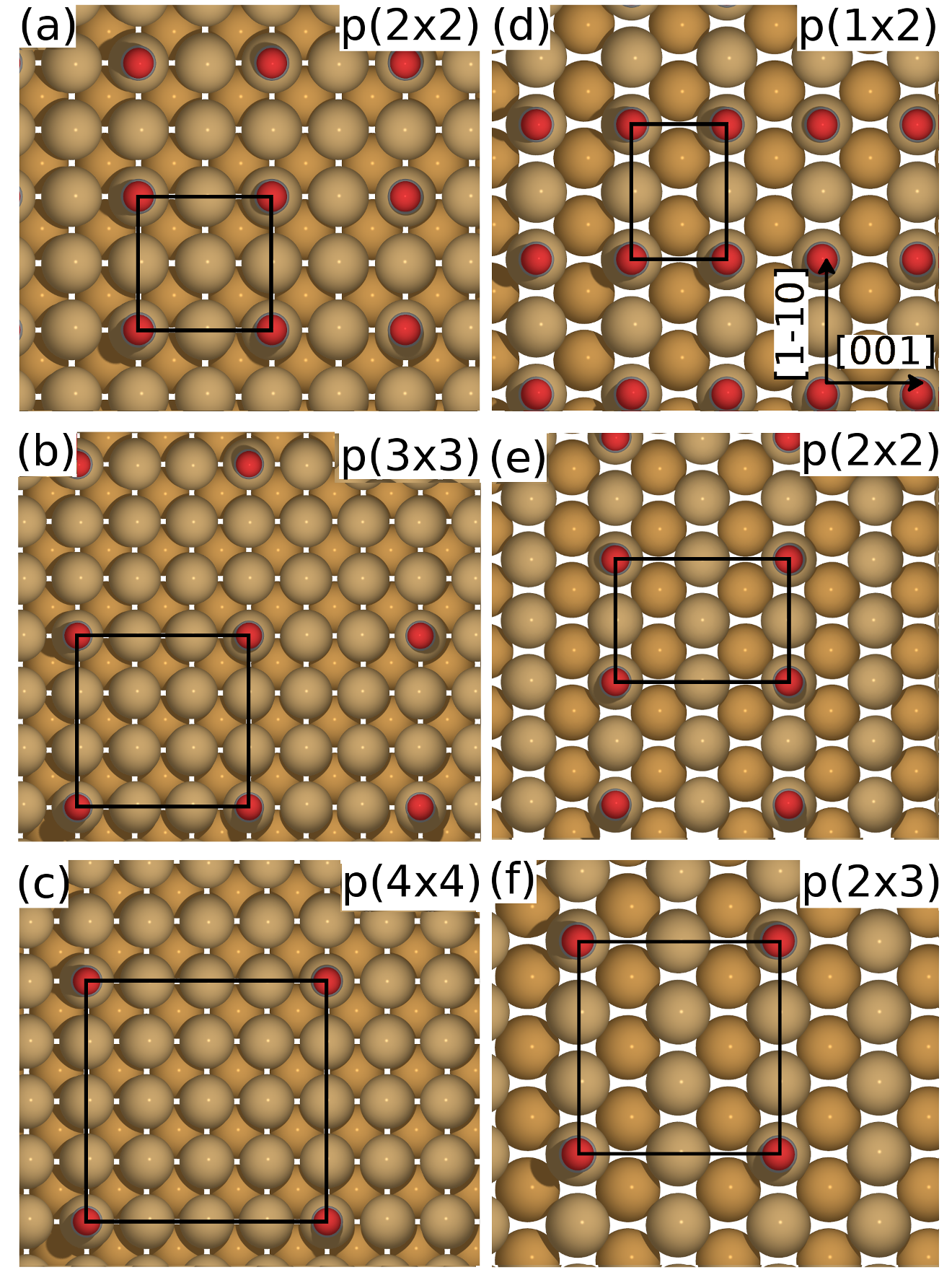}
\caption{\label{fig-diatomics-structure} (a)--(c) Top view of CO adsorbed on Cu(100) in a p(2x2) (a), a p(3x3) (b), and a p(4x4) overlayer (c).  (d)--(f) Top view of CO adsorbed on Cu(110) in a p(1x2) (d), a p(2x2) (e), and a p(2x3) overlayer (f). Odd and even layers of substrate are shown in different color tones.}
\end{figure}

We now show the utility of the electronic friction tensor to calculate EHP-induced nonadiabatic vibrational lifetimes. 
Considering the wealth of literature on this subject, we will focus on comparison to existing references from simulation~\cite{Butler1979, Persson1980, Head-Gordon1992a, Head-Gordon1992, Hellsing1984, Krishna2006, Forsblom2007, Rittmeyer2015} and experiment.~\cite{Beckerle1991, Morin1992, Owrutsky1992, Hirschmugl1994,Germer1994, Lane2007, Omiya2014} The results reported in this section are obtained by projection of the electronic friction tensor on calculated vibrational normal mode coordinates using eq.~\ref{eq-vibrelax} (see supplemental material for corresponding frequencies~\cite{supplemental}). We study CO, NO, and CN adsorbed in their respective atop equilibrium positions for different coverages and surface facets of transition metals (see also Fig.~\ref{fig-diatomics-structure}).

\begin{table}
\caption{\label{tab-diatomics} Vibrational lifetimes of diatomic molecules on different metal surfaces projected from the friction tensor along vibrational normal modes using eq.~\ref{eq-vibrelax}. IS: internal stretch mode, SA: surface-adsorbate stretch mode, FR: frustrated rotation mode, FT: frustrated translation mode (see also Figs.~\ref{fig-vibmodes} and \ref{fig-diatomics-structure}). For the CO/Cu(110) systems the FT and FR modes are nonequivalent and given for both modes in the order: parallel and perpendicular with respect to the (110) surface ridges.}
\begin{tabular}{ccC{1.0cm}C{1.0cm}C{1.3cm}C{1.3cm}} \hline \noalign{\vskip 2pt} 
coverage & system & IS  & SA  & FR  & FT \\
   &     & \multicolumn{4}{c}{lifetime in ps} \\ \hline \noalign{\vskip 2pt} 
c(2x2) & CO/Cu(100) & 10.0  & 29.8  & 3.70  & 202  \\
p(2x2) & CO/Cu(100) & 5.84  & 24.3  & 4.28  & 202  \\
p(3x3) & CO/Cu(100) & 4.53  & 40.6  & 3.87  & 202  \\ 
p(4x4) & CO/Cu(100) & 3.20  & 20.0  & 2.60  & 125  \\ \hline \noalign{\vskip 2pt} 
Ref. \onlinecite{Head-Gordon1992} & CO/Cu(100) & 3.3  & 82   & 2.3  & 108 \\
Ref. \onlinecite{Krishna2006}     & CO/Cu(100) & 3.3  & 13.7 & 3.8  & 19.5   \\ \hline \noalign{\vskip 2pt} 
p(2x1) & CO/Cu(110)  &   12.3  &  78.5  &  5.77/9.43     &  1345/943   \\
p(2x2) & CO/Cu(110)  &   3.61  &  28.3  &  3.27/4.47     &  168/303   \\
p(2x3) & CO/Cu(110)  &   3.36  &  25.0  &  4.25/4.21     &  251/317   \\
\multirow{10}{*}{\begin{sideways}p(2x2)\end{sideways}} 
 & CO/Cu(111) &  5.90 & 30.1  & 4.45  & 197 \\   
 & CO/Ag(111) &  4.11 & 24.0  & 4.92  & 1490 \\
 & CO/Au(111) &  4.83 & 15.6  & 4.26  & 2290 \\
 & CO/Ni(111) &  3.29 & 6.82  & 3.10  & 68.0 \\
 & CO/Pt(100) &  4.84 & 10.1  & 4.33  & 107 \\   
 & CO/Pt(111) &  2.77 & 5.52  & 4.90  & 146 \\ 
 & NO/Pt(111) &  1.32 & 2.58  & 0.21  & 13.3 \\
 & CN/Pt(111) &  2.53 & 18.2  & 8.30  & 79.1 \\
 & CN/Cu(111) &  36.5 & 48.1  & 6.64  & 37.2 \\
 & CN/Ag(111) &  111  & 62.3  & 8.10  & 47.8 \\  \hline \noalign{\vskip 2pt} 
Ref. \onlinecite{Forsblom2007} & CO/Pt(111) & 1.8 & - & - & - \\
Ref. \onlinecite{Krishna2006}  & CO/Pt(111) & 5.0 & 11.7 & 7.3 & 29.4 \\ 
Ref. \onlinecite{Forsblom2007} & CN/Pt(111) & 2.4 & - & - & - \\ 
Ref. \onlinecite{Krishna2006}  & CN/Pt(111) & 15.0 & - & 15.5 & - \\
Ref. \onlinecite{Forsblom2007}  & CN/Cu(111) & 22 & - & - & - \\
Ref. \onlinecite{Forsblom2007}  & CN/Ag(111) & 39 & - & - & - \\
Ref. \onlinecite{Krishna2006}  & NO/Pt(111) & 8.2 & 9.6 & 8.3 & 1.3 \\ 
 \hline
\end{tabular} 
\end{table}

EHP-mediated vibrational relaxation is the dominant energy loss mechanism of high-frequency adsorbate vibrations, decoupled from  substrate phonons. In such cases nonadiabatic energy loss can be measured by pump-probe spectroscopy~\cite{Beckerle1991, Morin1992} and indirectly by using substrate laser-heating.~\cite{Germer1994,Lane2007} One of the most studied systems is the internal stretch (IS) motion of CO adsorbed on Cu(100).~\cite{Morin1992, Harris1990, Germer1994, Ryberg1985a, Head-Gordon1992} The vibrational frequencies of CO have been measured to be  2078~cm$^{-1}$ (257.8~meV) for the internal stretch (IS) mode,~\cite{Hirschmugl1994} 345~cm$^{-1}$ (42.8~meV)~\cite{Hirschmugl1994} for the surface-adsorbate stretch (SA), 288~cm$^{-1}$ (35.3~meV)~\cite{Hirschmugl1994} for the frustrated rotation (FR) and 32~cm$^{-1}$(4~meV)~\cite{Graham2003} the frustrated translation (FT) (see also Fig.~\ref{fig-vibmodes}). Experimentally, \citet{Morin1992} determined the  vibrational lifetime for the IS mode to be 2$\pm$1~ps. Using infrared reflection absorption spectroscopy \citet{Hirschmugl1990} determined a lower bound for the FR lifetime as $\ge$1~ps. \citet{Persson1991} estimated that frustrated translation of CO on Cu has an EHP-induced lifetime of 39~ps.~\cite{Persson1991} Persson and Persson derived the lifetime of the IS mode as 1.8~ps by relating the relaxation rate with the charge transfer per oscillation period.~\cite{Persson1980}

We have recomputed the vibrational lifetimes of CO on Cu(100), as previously calculated by \citet{Head-Gordon1992} and \citet{Krishna2006} at different coverages (See Table~\ref{tab-diatomics}). We find our current approach to be in good agreement with the previous calculations performed in local basis representation using Hartree-Fock and small cluster models~\cite{Head-Gordon1992} and performed in a plane-wave implementation.~\cite{Krishna2006} As already previously reported,~\cite{Askerka2016} our calculations support the experimental findings of highly mode-dependent vibrational lifetimes and an overall faster decay following FR excitation than IS. The benefit of our current setup is that all vibrational lifetimes can be calculated using a single vibrational normal mode calculation. We have recently reported on the possibility of intermode coupling mediated by the fact that the Hessian and friction tensor do not commute.~\footnote{The here reported results for p(2x2) CO/Cu(100) differ slightly from previously reported ones,~\cite{Askerka2016} simply due to additional surface relaxation effects accounted for in the current data.} This would amount to electronic friction-induced coupling between modes that adds to already existing anharmonicity-imposed mode coupling. With the ability to calculate the full friction tensor we can study these couplings by comparison to vibrational lifetimes along Hessian normal modes and along eigenmodes of the friction tensor (see Table SII in the supplemental material~\cite{supplemental}). The larger the discrepancy in projected lifetime, the larger is the mode coupling introduced by tensorial electronic friction. For example in the case of CO/Cu(100), we find that IS and SA lifetimes extracted from friction modes (4.91~ps and 111~ps for p(2x2)) are significantly modified compared to IS and SA lifetimes calculated along the vibrational normal modes (5.84~ps and 24.3~ps for p(2x2)), whereas FR lifetimes are almost identical (4.21~ps vs. 4.28~ps). Therefore friction-induced mode coupling enhances energy loss along the IS mode and reduces energy loss along the SA mode.

For the high-frequency IS mode we find lifetimes of 3.20--10.0~ps for different coverages. This compares to an experimental lifetime of 2$\pm$1~ps measured at a c(2x2) coverage.~\cite{Morin1992} This discrepancy will be discussed further below. Low-frequency modes , such as FR and SA, are expected to dominantly lose energy through vibrational coupling and phonon dissipation.~\cite{Tully1993} In both cases, our EHP-mediated lifetimes serve as upper bounds to experiment. It is interesting to note that the IS lifetime extracted from a single CO adsorbed at the (7x7) Cu(100) terminated cluster is in closer agreement with experiment (2.75~ps). 

Comparing CO/Cu(100) lifetimes as a function of coverage, we find that overall lifetimes are larger for higher coverages. Adsorbate-adsorbate interactions and changes in the substrate DOS are stronger at higher coverages and may potentially reduce nonadiabatic effects along a single mode (here $\mathbf{q}=0$). However,  by only calculating lifetimes along $\mathbf{q}=0$ using eq.~\ref{eq-vibrelax} rather than eq.~\ref{eq-vibrelax2}, we neglect dephasing and mode coupling between adsorbates. Such coupling may lead to population of adsorbate modes with $\mathbf{q}\neq0$. In section~\ref{results-c2x2}, we will show that $\mathbf{q}\neq0$ modes can have significantly shorter lifetimes. We therefore expect $\mathbf{q}=0$ lifetimes at high coverage to misrepresent experimentally observed lifetimes. The coverage dependence of nonadiabatic adsorbate/substrate energy-transfer has been studied experimentally for the IS mode of CO on Cu(110)~\cite{Omiya2014} and CO on Cu(111).~\cite{Owrutsky1992} \citet{Owrutsky1992} find SFG linewidths at 0.10 monolayer (ML) and 0.45~ML that correspond to 1.6 and 2.5~ps, which qualitatively agrees with the trend we observe for CO on Cu(100) and Cu(110). However, experimental linewidths are governed by many different contributions and direct comparison with vibrational lifetimes is difficult. In contrast, \citet{Omiya2014} find a  faster vibronic coupling response of CO on Cu(110) to laser heating of the substrate at higher coverage. \citet{Springer1994,Kindt1998}, like others before, argue that nonadiabatic energy transfer from the substrate to the molecule will dominantly occur via excitation of the FR mode due to its high nonadiabatic relaxation rate (low vibrational lifetime) and its frequency being close to the substrate phonon frequencies. Comparing FR lifetimes of CO/Cu(110) at different coverages, we do not find a conclusive trend with coverage and different lifetimes along the [1-10] (parallel to ridges) and [001] (perpendicular to ridges) directions. However, the measured time-resolved change in SFG linewidth of CO on Cu(110) in Ref.~\onlinecite{Omiya2014} does not show a monotonic trend. Unfortunately, a direct comparison of calculated lifetimes with transient spectroscopy and substrate-heating experiments is currently not possible. Therefore, explicit MDEF simulations of the sub-picosecond hot-electron induced dynamics similar to the desorption dynamics of Ref. \onlinecite{Springer1994} would be necessary.

\citet{Persson1980}, \citet{Forsblom2007}, \citet{Krishna2006}, and \citet{Tully1993} have discussed the strong dependence of vibrational lifetimes on the chemical composition of adsorbate and substrate as well as changes due to differences in substrate-adsorbate interactions. From our results in Table~\ref{tab-diatomics}, we can add to these findings the additional aspect of dependence on the surface facet for CO on copper surfaces. For comparable surface coverage, we find the following trend for the IS and FR modes: Cu(110)$>$Cu(100)$>$Cu(111). However, the lifetimes only vary between 3.61 and 5.90~ps for the IS mode and 3.27 and 4.45~ps for the FR mode between the different substrate facets. 
Comparing substrates of different reactivity and adsorbate-substrate interaction, no clear trend emerges for the IS and FR modes. However, the SA and FT modes, which most strongly depend on the covalent interaction strength between CO and the substrate show clear trends across the coinage metals (Au, Ag, Cu) to more reactive surfaces such as Ni(111) and Pt(111). In the case of Pt and Ni surfaces, the SA mode lifetime has similar magnitude as IS and FR, whereas for the coinage metals FR and SA differ by a factor of 3-7. \citet{Forsblom2007} have interpreted these trends as coupled to the position of the acceptor level in the molecule with the Fermi level. Molecular resonances close to the Fermi level increase the density of state and the number of possible excitations at low frequency in eq.~\ref{eq-Gamma}. 

Next, we study trends between different adsorbates, including CO, CN, and NO on Pt(111). We find similar IS lifetimes for all three adsorbates ranging from 1.32 for NO to 2.77~ps for CO. The latter is in fair agreement with the experimentally measured lifetime of CO/Pt(111) of 1.8~ps.~\cite{Beckerle1991} However, all other modes show strong variations. The strongly hybridized NO molecule exhibits fast EHP-mediated energy dissipation along all modes, whereas CO and CN show longer lifetimes for the FT mode and in the case of CN also for the SA mode. \citet{Peremans1995} have measured the vibrational lifetime of the IS mode for CO/Pt(100) to be 1.5$\pm$0.5~ps using SFG and single crystal electrodes in aqueous electrolytes, whereas we find a lifetime of 4.84~ps. \citet{Matranga2000} have measured the IS mode lifetime of CN on Pt(111) and a polycrystalline silver interface using SFG and found potential-dependent lifetimes between 3 and 8~ps and 28 and 60~ps. These measurements agree in orders of magnitude with our calculated lifetimes of 2.77 and 111~ps. We find surprisingly low vibrational lifetimes for NO on Pt(111) with 1.32~ps for the IS mode and 0.21~ps for the FR mode. The vibrational relaxation of NO is known to be fast. For NO on Ir(111) the frustrated rotation is found to  respond to hot electrons at a time scale of 700~fs.~\cite{Lane2007}  In agreement with our calculations, \citet{Abe2003} have estimated the NO on Pt(111) IS mode lifetime to be roughly 4 times shorter than the CO on Cu(111) lifetime, which when applied to our case would amount to 1.4~ps. We find 1.32~ps. 

We find some discrepancies with previously published results from \citet{Krishna2006} and \citet{Forsblom2007} (shown in Table~\ref{tab-diatomics}) which may be due to the different choice of exchange-correlation functional approximation (PBE and PW91~\cite{Perdew1992}), the different choice of relaxation rate expression (eq.~\ref{eq-appendix-gamma-dos4} vs. eq.~\ref{eq-appendix-double-delta}), and the different choice of broadening method (gaussian and square window). Using the previously employed~\cite{Krishna2006} square window method of broadening we found results to be highly sensitive on the employed broadening width $\sigma$, which complicates the calculation of stable relaxation rates. Overall we find our calculations to be in better agreement with results of \citet{Forsblom2007}, which have been obtained from plane-wave calculations using eq.~\ref{eq-appendix-double-delta} and the Perdew-Wang exchange correlation functional.~\cite{Perdew1992} The case of NO on Pt(111) deserves some additional attention. Contrary to \citet{Krishna2006} where the FT mode is identified as the fastest dissipation channel (1.3~ps), we find the IS and FR lifetimes to be the fastest. We found this result to  hold also for a larger number of explicitly treated substrate layers and for different xc-functionals.

We conclude this section with an analysis of the relevance of off-diagonal elements in the electronic friction tensor. As already discussed above, the off-diagonal elements mediate nonadiabatic energy transfer between degrees of freedom. Therefore, when neglecting off-diagonal elements, differences in relaxation rates along vibrational modes become smaller. For example, for p(2x2) CO on Cu(100), the lifetimes along the IS mode and the SA mode are 8.25 and 11.0~ps instead of 5.84 and 24.3~ps. Equally, the FT lifetime is largely overestimated and changes from 202~ps to 15.6~ps. The importance of off-diagonal elements in the friction tensor for molecular adsorbates is not so much surprising, considering that also the dynamical matrix contains off-diagonals in Cartesian representation. An important question is if such off-diagonal elements in the tensor are also significant in the normal mode representation (see eq.~\ref{eq-normalmoderepresentation}), where the diagonal elements of the friction tensor correspond to the lifetimes presented in Table~\ref{tab-diatomics}. In the case of CO on Cu(100) in its equilibrium position we find most off-diagonals to be close to zero, however, significant couplings exist between FT and FR modes ($\pm$0.03~ps$^{-1}$) and between SA and IS modes (-0.07~ps$^{-1}$). These correspond to changes in lifetime on the order of tens of ps. While they may not seem significant in this particular case, their magnitude may be larger and non-negligible at nonequilibrium positions and during explicit dynamics.

\subsection{Electronic friction and intermolecular coupling - c(2x2) CO on Cu(100)}
\label{results-c2x2}

\begin{figure}
\centering\includegraphics[width=\columnwidth]{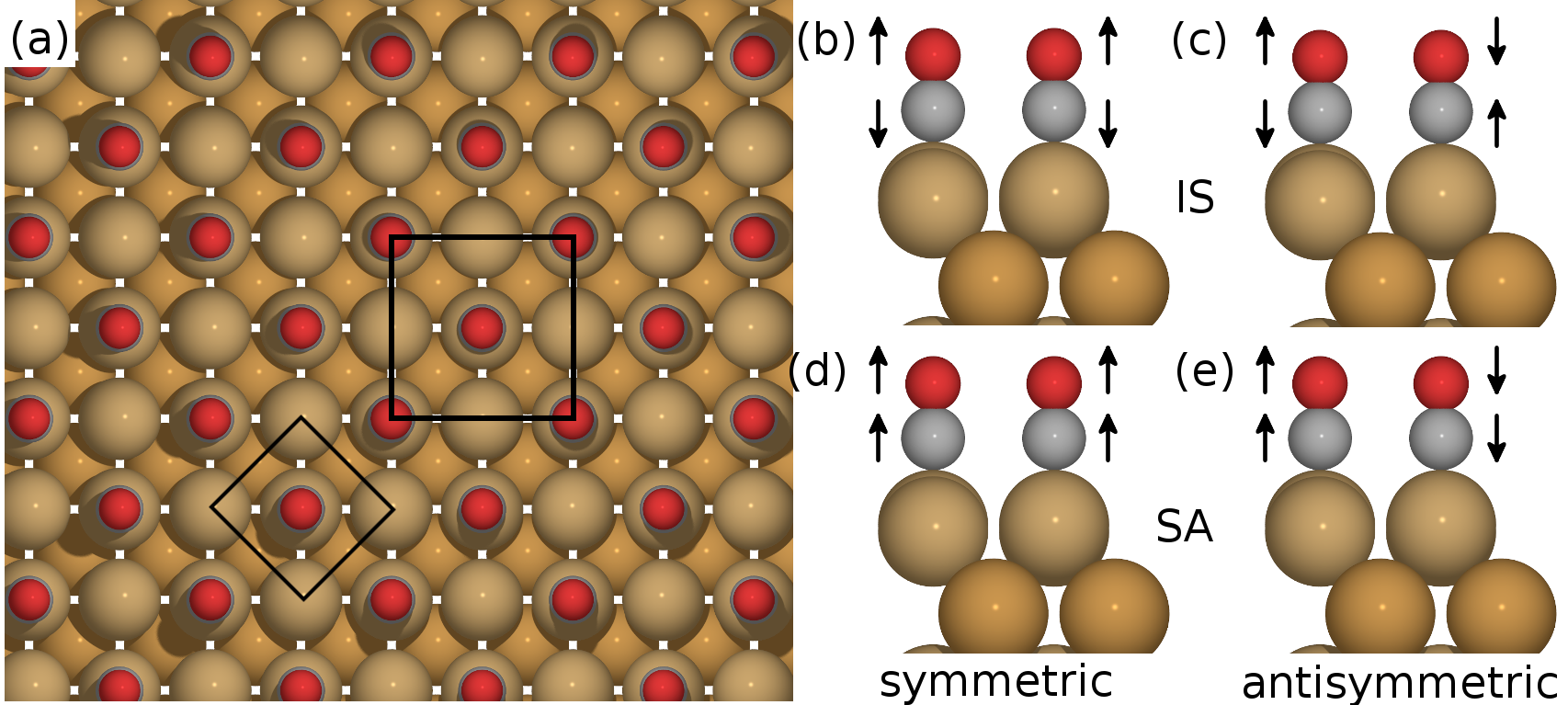}
\caption{\label{fig-c2x2} (a) Top view of c(2x2) CO on Cu(100) overlayer. Shown are also two different equivalent unit cells each containing one or two molecules. (b) -- (e) Symmetric and antisymmetric adsorbate phonon combinations for internal stretch (IS) and surface-adsorbate stretch (SA). Odd and even layers of substrate are shown in different color tones.}
\end{figure}

In this section, we discuss electronic friction-induced coupling between neighboring adsorbate molecules. Up to this point, we have studied surface unit cells featuring only one adsorbate molecule. Therefore, we have only described totally symmetric ($\mathbf{q}$=0) collective motion of the adsorbate overlayer (see Fig.~\ref{fig-c2x2}(b) and (c)). However, this coherent motion can dephase into other wave vector modes, due to scattering with low-frequency phonons and, as we will see, due to scattering with electron-hole pairs. This dephasing often occurs on a time scale similar as or faster than vibrational relaxation.~\cite{Matsumoto2006} In the case of CO on Cu(100), \citet{Morin1992} report a line width that corresponds to an overall dephasing time of 1.2~ps, and a electron-hole pair dominated relaxation time of 2~ps. If we attribute the difference to pure dephasing, we arrive at a decoherence time of 3~ps.

This means that, during vibrational relaxation initially excited collective adsorbate motion will dephase into different linear combinations of adsorbate modes that can be described through a Fourier expansion and a wave vector $\mathbf{q}$ within the first Brillouin zone. For example the totally symmetric and totally antisymmetric adsorbate phonons correspond to wave vectors pointing to the center ($\Gamma$ point) and the edge of the Brillouin zone ($M$ point for fcc(100) surface unit cell). We can equally model an antisymmetric linear combination of vibrational modes in a unit cell featuring two molecules, such as in a p(2x2) unit cell. This effectively corresponds to a c(2x2) arrangement described in a non-primitive unit cell (see Fig.~\ref{fig-c2x2}(a)). We have calculated the (12$\times$12) electronic friction tensor for this overlayer in equilibrium position using a 20$\times$20$\times$1 $\mathbf{k}$-grid and otherwise the same settings as outlined above.


\begin{table}
\caption{\label{tab-c2x2} Frequencies and lifetimes of vibrational modes for 2 CO molecules  in a c(2x2) overlayer on Cu(100).}
\begin{tabular}{ccC{2.1cm}C{2.1cm}} \hline\noalign{\vskip 2pt} 
\multicolumn{2}{c}{mode}  & frequency  & lifetime \\ 
 &  & in cm$^{-1}$ &  in ps \\ \hline \noalign{\vskip 2pt} 
symm. &FT  & 31  & 202  \\ 
asymm. &FT & 20  & 287  \\
average &FT & - & 237 \\
symm. &FR  & 280 & 3.80 \\
asymm. &FR & 275 & 1.66 \\
average &FR & - & 2.31 \\
symm. &SA  & 328 & 29.1 \\
asymm. &SA & 327 & 6.74 \\
average &SA & - & 11.2 \\
symm. &IS  & 2059& 10.0 \\ 
asymm. &IS & 1974& 3.30 \\ 
average &IS & - & 4.96 \\ \hline \noalign{\vskip 2pt} 
\end{tabular} 
\end{table}

In the redundant unit cell, we can model symmetric and antisymmetric combinations, in total 12, of the previously discussed adsorbate vibrations. To our great surprise, we find large differences in EHP-induced vibrational lifetimes between them (see Table~\ref{tab-c2x2}). The symmetric combinations, as expected,  correspond to the same lifetimes we can describe in the primitive unit cell (see Table~\ref{tab-diatomics}). The antisymmetric combination modes, with the exception of the FT mode, show significantly shorter lifetimes. This difference is especially drastic for the IS and SA modes. In the case of the FR and SA modes, the lifetime changes from 3.80 to 1.66~ps and 29.1 to 6.74~ps despite a negligible difference in vibrational frequency between symmetric and antisymmetric adsorbate phonons.

The calculated differences in lifetimes are rooted in the off-diagonal elements of the electronic friction tensor responsible for coupling  different CO molecules. In the supplemental material, we present the (12$\times$12) electronic friction tensor in both Cartesian and normal mode representations as well as its eigenvalues.~\cite{supplemental} In Cartesian representation, we find significant off-diagonal elements between atoms of different molecules. The lifetimes we report in Table~\ref{tab-c2x2} are the inverse of the diagonal elements of the friction tensor in the normal mode representation. In that representation, the friction tensor shows a block-diagonal structure, which is governed by the symmetry of the equilibrium site. Modes perpendicular to the surface do not couple with those that are parallel to the surface. Even in the normal mode representation, we find significant off-diagonal elements, albeit smaller than in Cartesian representation. We can therefore add to our previous finding of friction-induced mode coupling between intramolecular modes the fact that friction also induces coupling between neighboring adsorbate molecules. This electron-hole pair mediated coupling will contribute to the dephasing of collective overlayer motion.

In Table~\ref{tab-diatomics} we show a trend of larger lifetimes for $\mathbf{q}=0$ motion at higher coverages, amounting to 10~ps for the IS mode at a c(2x2) coverage. However, in the case of CO on Cu(100) at a c(2x2) coverage, a lifetime of 2$\pm$1~ps has been observed for the transient vibrational damping of high coverage overlayers.~\cite{Morin1992} Comparing experiment with our calculated $\mathbf{q}=0$ relaxation rate of 10~ps would assume that no dephasing occurs over the time span of the relaxation. This assumption does not correctly reproduce the experimentally observed lifetime nor the coverage dependence infered by other works~\cite{Omiya2014}. The other limiting case would be to assume rapid dephasing, which corresponds to calculating the relaxation rate of individual decoherent CO molecules as an average of different wave vectors following eq.~\ref{eq-vibrelax2}. In the case of the  IS mode the antisymmetric mode corresponds to 3.3~ps rather than 10.0~ps for the symmetric combination. Their average is 4.96~ps, which is in closer agreement with experiment. However, both of these assumptions are limiting cases and the real combination of phonon modes contributing to relaxation will change within the time span of relaxation, which is a picture that cannot be represented by static calculation of relaxation rates. Modern high resolution experiments might be able to identify this process as non-exponential decay and a change of relaxation rate over time. It should furthermore be noted that the corresponding friction eigenvalue, which is a combination of symmetric and antisymmetric SA and IS modes only corresponds to a lifetime of 2.5~ps. In summary, we find that the tensorial structure of the friction tensor is important for an accurate description of experimentally observed population decay at high surface coverages, which is governed by an interplay of vibrational relaxation and decoherence dephasing. Further work, analyzing in detail the coupling contributions and the role of decoherence dephasing, is needed.


\section{Conclusions and Outlook}
\label{conclusions}

Nonadiabatic energy transfer is an important aspect of dynamics at metal surfaces. Electronic excitations in the substrate partly control adsorbate dynamics and reduce vibrational lifetimes. In this work, we have revisited an approach to calculate nonadiabatic vibrational energy loss of atoms and molecules at metal surfaces based on time dependent perturbation theory. We have presented an \emph{ab-initio} implementation and calculations of the electronic friction tensor in the zero frequency limit, using Density Functional Theory in a local basis representation.  We have given a detailed discussion of the numerical stability and convergence properties when using broadening-based Brillouin zone integration methods. Common approximations for the coupling matrix elements do not introduce significant deviations for periodic systems, however may yield deviations for smaller metal clusters. The calculation of EHP-induced vibrational lifetimes for diatomic and multiatomic adsorbates is feasible at the cost of a harmonic normal mode calculation. Our results are largely  consistent with previous calculations stemming from cluster models and from plane-wave based periodic methods, despite some deviations that arise due to different numerical settings. Comparison to experimental data shows agreement within the same order of magnitude, but does not enable definitive conclusions on the systematic accuracy of the method. An interesting finding is the intermolecular coupling induced by the tensorial structure of electronic friction, which is critical to understand experimental observations. Nevertheless, explicit molecular dynamics simulations of dynamical observables and time-resolved spectroscopic signatures beyond the harmonic regime are still  necessary to assess the reliability of this approach and the importance of the tensorial description of electronic friction.

Our computational method can easily be combined with coupling matrix elements from Density Functional Perturbation Theory (Coupled Perturbed Kohn-Sham)~\cite{Trail2001, Quong1992, Liu1996, Savrasov1996, Baroni2001} and implemented in on-the-fly dynamics simulations. Considering approximate friction methods, such as the LDFA,~\cite{Juaristi2008,Blanco-Rey2014,Rittmeyer2015} this approach may serve as a reference that guides the improvement of LDFA beyond the local density limit, for example towards incorporation of semi-local contributions that correlate with the electron density gradient.  Future work would be needed to study the applicability and usefulness of our approach in nonadiabatic dynamics on surfaces and its potential extensions towards dynamics beyond the weak coupling limit,~\cite{Dou2016a} dynamics including explicit memory effects,~\cite{Olsen2010} and the efficient incorporation of excited state screening.~\cite{Maurer2013,Evangelista2013}

\begin{acknowledgments}
This research was supported primarily by the US Department of Energy - Basic Energy Science grant no. DE-FG02-05ER15677, with additional support for VSB provided by the Air Force Office of Scientific research grant no. FA9550-13-1-0020. Computational support was provided by the HPC facilities operated by, and the staff of, the Yale Center for Research Computing. MA thanks Christian Negre for helpful discussions. RJM acknowledges fruitful discussions with Volker Blum during a visit at Duke University.
\end{acknowledgments}

\appendix

\section{Two different expressions for vibrational damping rates in the quasi-static limit}
\label{appendix-alternative-expression}

Starting from 
\begin{align}\label{eq-app}
 \Gamma(\omega_{j}) &=\frac{\pi\hbar^2\omega_{j}}{M} \sum_{\nu,\nu'>\nu} 
    |g_{\nu,\nu'}^{j}|^2\cdot \\ \nonumber & 
  [f(\epsilon_{\nu})-f(\epsilon_{\nu'})]\cdot \delta(\epsilon_{\nu'}-\epsilon_{\nu}-\hbar\omega_{j}),
\end{align}
we explicitly introduce the electronic DOS $\rho(\epsilon)$,
\begin{equation}
 \rho(\epsilon) =  \sum_{\nu} \delta(\epsilon - \epsilon_{\nu}),
\end{equation}
with
\begin{equation}
 \int_{-\infty}^{\infty} d{\epsilon}\cdot \rho(\epsilon) =1,
\end{equation}
and arrive at
\begin{align}\label{eq-appendix-gamma-dos}
 \Gamma(\omega) &=\frac{\pi\hbar}{M} \int_{-\infty}^{\infty} d\epsilon_i \int_{-\infty}^{\infty} d\epsilon_f  \sum_{\nu,\nu'>\nu}  |g_{\nu',\nu}|^2 \cdot \\ \nonumber 
 & [f(\epsilon_{\nu})-f(\epsilon_{\nu'})] \cdot(\epsilon_{\nu'}-\epsilon_{\nu})\cdot \\ \nonumber 
 &\delta(\epsilon_f-\epsilon_{\nu'})\delta(\epsilon_i-\epsilon_\nu) \delta(\epsilon_f-\epsilon_i-\hbar\omega) .
\end{align}
In eq.~\ref{eq-appendix-gamma-dos} energies with indices $i$ and $f$ are continuous variables, whereas energies with indices $\nu$ and $\nu'$ are discrete variables. We now perform a variable change to $x=\epsilon_f-\epsilon_i$ and $y=\epsilon_f+\epsilon_i$:
\begin{align}\label{eq-appendix-gamma-dos2}
 \Gamma &=\frac{\pi\hbar}{M} \int_{0}^{\infty} dx \int_{-\infty}^{\infty} dy  \sum_{\nu,\nu'>\nu}  |g_{\nu',\nu}|^2 \cdot \\ \nonumber 
 & [f(\epsilon_{\nu})-f(\epsilon_{\nu'})] \cdot(\epsilon_{\nu'}-\epsilon_{\nu})\cdot \\ \nonumber 
 &\delta(y-\epsilon_i-\epsilon_{\nu'})\delta(y-\epsilon_f-\epsilon_\nu) \delta(x-\hbar\omega) .
\end{align}
By applying $\int dy \delta(x-y)\delta(y-z)=\delta(x-z)$ we find
\begin{align}\label{eq-appendix-gamma-dos3}
 \Gamma(\omega) &=\frac{\pi\hbar}{M} \int_{0}^{\infty} dx \sum_{\nu,\nu'>\nu}  |g_{\nu',\nu}|^2 \cdot 
  [f(\epsilon_{\nu})-f(\epsilon_{\nu'})] \cdot(\epsilon_{\nu'}-\epsilon_{\nu})\cdot \\ \nonumber 
 &\delta((\epsilon_{\nu'}-\epsilon_{\nu})-x) \cdot \delta(x-\hbar\omega) .
\end{align}
We can define the electron-phonon spectral function $A(x)$
\begin{equation}\label{eq-app-spectral-func}
 A(x) = \sum_{\nu,\nu'>\nu}  |g_{\nu',\nu}|^2 \cdot 
  [f(\epsilon_{\nu})-f(\epsilon_{\nu'})] \cdot(\epsilon_{\nu'}-\epsilon_{\nu})\cdot \delta((\epsilon_{\nu'}-\epsilon_{\nu})-x)
\end{equation}
which results in the following expression for the relaxation rate:
\begin{align}\label{eq-appendix-gamma-dos4}
 \Gamma(\omega) =\frac{\pi\hbar}{M} \int_{0}^{\infty} dx A(x) \cdot \delta(x-\hbar\omega) = \frac{\pi\hbar}{M} \cdot A(\hbar\omega) .
\end{align}
The quasi-static limit assumes that $\hbar \omega_j$ is much smaller than energy $x$ at which $A(x)$ starts changing from its low-energy asymptotic value, so $A(x)$ can be evaluated at $x=0$.


In the quasi-static limit we can also derive an alternative expression to eq.~\ref{eq-appendix-gamma-dos4} for nonadiabatic vibrational damping~\cite{Allen1972,Head-Gordon1992a} valid in the low temperature limit. Starting from eq.~\ref{eq-appendix-gamma-dos}, we change variables to $x=\epsilon_f-\epsilon_F$ and $y=\epsilon_i-\epsilon_F$ by shifting energies with respect to the Fermi level $\epsilon_F$ and assume 0~K electronic temperature:
\begin{align}
  \Gamma(\omega) &=\frac{\pi\hbar}{M} \int_{-\infty}^{0} dy \int_{0}^{\infty} dx  \sum_{\nu<\epsilon_F}\sum_{\nu'>\epsilon_F}  |g_{\nu',\nu}|^2 \cdot (\epsilon_{\nu'}-\epsilon_{\nu}) \cdot \\ \nonumber 
 &\delta(x+\epsilon_F-\epsilon_{\nu'})\delta(y+\epsilon_F-\epsilon_\nu) \delta(x-y-\hbar\omega) .
\end{align}
By applying $\int dy \delta(x-y)\delta(y-z)=\delta(x-z)$ again we find
\begin{align}\label{eq-appendix-double-delta2}
\Gamma(\omega) &=\frac{\pi\hbar}{M} \int_{0}^{\hbar\omega} dx \sum_{\nu<\epsilon_F}\sum_{\nu'>\epsilon_F}  |g_{\nu',\nu}|^2 \cdot (\epsilon_{\nu'}-\epsilon_{\nu}) \cdot \\ \nonumber 
&\delta(x+\epsilon_F-\epsilon_{\nu'})\delta(x+\epsilon_F-\hbar\omega-\epsilon_\nu).
\end{align}
In eq.~\ref{eq-appendix-double-delta2} the upper limit of the integral is given by $\epsilon_\nu=x+\epsilon_F-\hbar\omega\leq\epsilon_F$. In assuming that the perturbation $\hbar\omega$ is small and that the integral does not strongly depend on $x$, we can move the double sum before the integral and evaluate it analytically to yield an additional factor of $\hbar\omega$. We arrive at:
\begin{align}\label{eq-appendix-double-delta}
\Gamma &=\frac{\pi\hbar}{M} \sum_{\nu<\epsilon_F}\sum_{\nu'>\epsilon_F}  |g_{\nu',\nu}|^2 \cdot (\epsilon_{\nu'}-\epsilon_{\nu})^2\cdot \\ \nonumber 
&\delta(\epsilon_F-\epsilon_{\nu'})\cdot \delta(\epsilon_F-\epsilon_\nu) .
\end{align}
This is the low-temperature expression as it has been used numerous times in literature.~\cite{Butler1979,Head-Gordon1992a,Forsblom2007} In this work we, however, use eq.~\ref{eq-appendix-gamma-dos4}.

%


\end{document}